\documentclass[a4paper,11pt]{article}
\usepackage{jheppub} 
\usepackage{lineno}
\usepackage{amsmath,amsfonts,amssymb,physics,slashed}
\usepackage{cancel}
\usepackage[capitalize]{cleveref}
\usepackage{xcolor}
\usepackage{comment}
\usepackage{subcaption}
\usepackage{graphicx}
\usepackage{svg}
\usepackage{float}
\usepackage[force]{feynmp-auto}

\def\iu{{\rm i}}
\def\e{{\rm e}}
\def\vEW{{\rm v}}
\def\phihat{\varphi}

\newcommand{\rp}[1]{\textcolor{blue}{[{\bf RP:} #1]}}

\preprint{CALT-TH/2024-035}
\title{The cosmology of ultralight scalar dark matter coupled to right-handed neutrinos}

\author[a]{Ryan Plestid.}
\author[a]{Sophia Tevosyan}

\affiliation[a]{Walter Burke Institute for Theoretical Physics, California Institute of Technology, Pasadena, CA 91125, USA}
\emailAdd{rplestid@caltech.edu}
\emailAdd{stevosya@caltech.edu}

\abstract{We consider ultralight scalar dark matter that couples to right-handed neutrinos. Due to the high density of neutrinos in the early universe, the background neutrino density dominates the dynamics of the scalar field, and qualitatively alters the field's cosmological evolution. This effect has not been included in previous literature, and changes the interpretation of cosmological data and its interplay with laboratory experiments. To illustrate these points a simplified model of a $1+1$ setup with a single scalar field is analyzed. 

We find that: {\it i}) The scalar field experiences an asymmetric potential and its energy density redshifts differently than ordinary matter. {\it ii}) Neutrino mass measurements at the CMB and oscillation experiments performed today complement one another (i.e., they constrain different regions of parameter space). {\it iii}) There exists potentially interesting cosmologies with either $O(1)$ variations in the dark matter density between the CMB and today, or $O(1)$ oscillations of neutrino mass. }

\begin{document}
\maketitle
\flushbottom

\vfill 

\pagebreak 

\section{Introduction \label{sec:intro}}
The identity of dark matter and the origin of neutrino masses are two of the most important outstanding questions in fundamental physics \cite{deGouvea:2016qpx,Cooley:2022ufh,Antypas:2022asj}. Most models that predict the origin of neutrino mass involve states that are singlets under the Standard Model gauge group (e.g., right-handed neutrinos) \cite{deGouvea:2016qpx}; the same is true of many dark matter models \cite{Cooley:2022ufh,Antypas:2022asj}. Since no gauge symmetry forbids their interaction, it is natural to consider couplings between dark matter and the progenitors of neutrino mass. In this paper we will be interested specifically in ultralight dark matter (ULDM with $m_\phi \ll 1~{\rm eV}$), which acts as a classical field and whose oscillations can imprint observable signals in the neutrino sector.

Due to quantum statistics and simple phase space considerations, ULDM must have integer spin and obey Bose statistics \cite{Ferreira:2020fam,Cheong:2024ose}. As a result, arguably the simplest model of ultralight dark matter (ULDM) is the oscillating homogeneous mode of a scalar field. This model of ULDM can easily accommodate the observed dark matter relic abundance via the so-called misalignment mechanism \cite{Preskill:1982cy, Abbott:1982af, Dine:1982ah, Turner:1983he}. As already emphasized, assuming that the scalar field is a Standard Model gauge singlet, it will generically couple to the right-handed neutrino mentioned above. It is therefore natural, in this context, to consider the interplay between dark matter and neutrino phenomenology. 

This simple observation has motivated a variety of studies involving neutrinos coupled to scalar fields \cite{Fardon:2003eh, Brookfield:2005bz, Kaplan:2004dq, Afshordi:2005ym, Barger:2005mn,Fardon:2005wc, Lee:1977ua, Krnjaic:2017zlz, Brdar:2017kbt, Berlin:2016woy, Huang:2022wmz, Whisnant:2018oft, Huang:2018cwo, Dev:2020kgz, Losada:2021bxx, Dev:2022bae, Murgui:2023kig, Blennow:2019fhy, Barman:2022scg, Farzan:2012sa, Ghalsasi:2016pcj,Ma:2006fn, Ma:2006km, Berlin:2016bdv, Capozzi:2018bps, Reynoso:2016hjr, Cline:2019seo, Choi:2019ixb, Huang:2021kam, Pandey:2018wvh, Chun:2021ief, Choi:2019zxy, Reynoso:2022vrn, Farzan:2018pnk, Farzan:2019yvo, Ge:2019tdi, Zhao:2017wmo,Davoudiasl:2023uiq,CarrilloGonzalez:2020oac, Martinez-Mirave:2024dmw,ChoeJo:2023ffp, Goertz:2024gzw}. This includes connections between mass-varying neutrinos and dark energy \cite{Fardon:2003eh, Brookfield:2005bz, Kaplan:2004dq, Fardon:2005wc,Afshordi:2005ym, Barger:2005mn, Lee:1977ua,Sakstein:2019fmf} and neutrinos coupled to bosonic dark matter \cite{Krnjaic:2017zlz, Brdar:2017kbt, Berlin:2016woy, Huang:2022wmz, Whisnant:2018oft, Huang:2018cwo, Dev:2020kgz, Losada:2021bxx, Dev:2022bae, Murgui:2023kig, Blennow:2019fhy, Barman:2022scg, Farzan:2012sa, Ma:2006fn, Ma:2006km, Berlin:2016bdv, Capozzi:2018bps, Reynoso:2016hjr, Cline:2019seo, Choi:2019ixb, Huang:2021kam, Pandey:2018wvh, Chun:2021ief, Choi:2019zxy, Reynoso:2022vrn, Farzan:2018pnk, Farzan:2019yvo, Ge:2019tdi, Zhao:2017wmo}. In the present context, where the scalar field is a dark matter candidate,  Refs.~\cite{Berlin:2016woy,Krnjaic:2017zlz, Brdar:2017kbt,   Whisnant:2018oft, Huang:2018cwo, Dev:2020kgz, Losada:2021bxx, Huang:2022wmz,Dev:2022bae} have considered the potentially observable signals of so-called ``distorted neutrino oscillations'' (DiNOs) which may be observed in terrestrial experiments. These DiNOs arise from time-dependent neutrino masses which appear from expanding the see-saw like formula, 
\begin{equation}
    m_\nu \sim  \frac{m_D^2}{m_N-g\phi(t)}\simeq \frac{m_D^2}{m_N}\qty(1+ \frac{g \phi(t)}{m_N} + \ldots)~, 
\end{equation}
where $m_N$ is the bare Majorana mass, $m_D$ is the Dirac mass, and $g$ is the scalar right-handed neutrino coupling constant. 

A common constraint that is discussed in the literature arises from the dynamics of $\phi$ in the early universe. If one assumes that the scalar field's energy density red-shifts like matter, then at earlier epochs the field's amplitude scales like $A_\phi \sim (1+z)^{3/2}$. If $ g A_\phi(z)/m_N \sim O(1)$ at the epoch of the cosmic microwave background (CMB), then the sum of neutrino masses, $\sum m_\nu$, can be used to set constraints on the model \cite{Berlin:2016woy, Krnjaic:2017zlz, Brdar:2017kbt}. 

Notice, however, that when $g A_\phi /m_N \sim O(1)$ the behavior of the system is very different than for $g A_\phi \ll m_N$. The left- and right-handed neutrinos will form a Dirac pair for $g \phi= m_N$, and transition from light, $\nu_L$,  to heavy, $\nu_H$, states. This can induce decays $\nu_H \rightarrow \nu_L \phi$. Furthermore, if $\phi$ crosses a critical value $\phi_c =m_N/g$, then all the light neutrinos in the bath will become heavy, and this incurs an enormous energetic cost. As we will see in what follows, this energetic cost dominates the behavior of the scalar field's potential in the early universe. This then substantially modifies the evolution of the scalar field, and the interpretation of constraints from $\sum m_\nu$ from the CMB. 

\pagebreak 

The purpose of this paper is to study the dynamics of a scalar field coupled to right-handed neutrinos, properly accounting for the energy density of the background neutrino  gas. This alters the scalar field's dynamics and depending on the initial conditions of the scalar field can lead to interesting phenomenology.

In what follows we will study a simplified model of neutrino scalar interactions. Our Lagrangian is defined by 
\begin{equation}
    \label{model-def}
    \mathcal{L} = \mathcal{L}_{\rm SM} + \frac{1}{2} \partial_\mu \phi \partial^\mu \phi + \bar{n}^c \iu \slashed{\partial} n^c - V_0(\phi)  - (g \phi - m_N) n^c n^c  - y L H n^c + \text{h.c.}~, 
\end{equation}
where $y$ is a Yukawa coupling, $L$ is the left-handed lepton doublet, $H$ is the Higgs doublet, $n^c$ is a right-handed neutrino, $\phi$ is the scalar field, $g$ the coupling constant, and $V_0(\phi)$ is the bare scalar potential. In what follows for simplicity, definiteness, and ease of comparison with other work, we will take 
\begin{equation}
\label{V_0}
    V_0(\phi) = \frac12 m_\phi^2 \phi^2~. 
\end{equation}
The scalar field $\phi$ is our dark matter candidate. When $\phi$ oscillates about the origin, $m_\phi$ is the mass of ULDM, and $m_\nu \sim m_D^2/m_N$ with $m_D = y \vEW/\sqrt{2}$, and $\vEW$ the Higgs vacuum expectation value. We will assume that the physical scalar mass, $m_\phi$, can assume very small values and do not consider naturalness issues related to the fine tuning of radiative corrections.

The rest of the paper is organized along the following lines. In \cref{sec:zero-cross} we discuss preliminaries that are necessary to understand the scalar field's dynamics. In \cref{scalar-dynamics} we discuss how the scalar field evolves throughout cosmic history. Next in \cref{sec:cosmo-const} we consider various cosmological scenarios related to initial conditions and the temperature at which oscillations begin. In \cref{sec:pheno} we discuss a variety of constraints from cosmology, astrophysics, and laboratory experiments. Finally in \cref{sec:conclusions} we summarize our findings and comment on potentially interesting future directions. 

\section{Zero crossings of the right-handed Majorana mass \label{sec:zero-cross} }
In this section we will motivate why zero crossings (where the Majorana mass of $n^c$ vanishes) appear when the dynamics of the scalar field are treated naively. This illustrates how and why the background potential from the neutrino gas must be included in the scalar dynamics. 

To see why zero-crossings are generic when considering ULDM, let us work backwards from the present day assuming a standard misalignment mechanism. The (galactic) energy density of dark matter today is known and given by \cite{Weber_2010,Planck:2018vyg}
\begin{equation}
    \rho_{\rm DM,now} \approx 9 \times 10^{-12}~{\rm eV}^4~. 
\end{equation}
Assuming the scalar field $\phi$ oscillates about the origin, this fixes  
\begin{equation}
  A_\phi(z=0) = \frac{\sqrt{2 \rho_{\rm DM,now}}}{m_\phi} \approx 4 \times 10^{10}~{\rm eV}  \qty(\frac{10^{-16}~\rm eV}{m_\phi})~. 
\end{equation}
In the standard misalignment mechanism, the field begins to oscillate at a temperature $T_{\rm osc}$ defined by $3 H(T_{\rm osc})= m_\phi$. Blue-shifting the field back to $T_{\rm osc}$ we find 
\begin{equation}
    A_\phi(z_{\rm osc}) \sim  \qty(\frac{T_{\rm osc}}{T_{\rm now}})^{3/2} A_\phi(z=0) \approx 3 \times 10^{24}~{\rm eV} \qty(\frac{0.65}{\gamma(T_{\rm osc})})^{3/4} \qty(\frac{10^{-16}~\rm eV}{m_\phi})^{1/4}~,
\end{equation}
where $H=\gamma T^2/M_P{\rm Pl}$ with $M_{\rm Pl}= 2.43\times 10^{18}~{\rm GeV}$ the reduced Planck mass, and $\gamma^2 = \pi^2 g_*(T)/90$ with $g_*(T)$ the effective number of degrees of freedom. We take as $g_*(T)\approx3.9$ for $T_{\rm osc}\approx0.1~ {\rm MeV}$. We see that even for rather small values of $m_\phi$ the amplitude of the field at the onset of oscillations is within three orders of magnitude of the Planck mass. It is therefore very difficult, without considering very small values of $g$, to avoid $A_\phi \simeq m_N/g$. 

As we will see in what follows, the dynamics of the system at large energy densities are not properly accounted for by simply re-scaling $A_\phi \sim (1+z)^3$. To understand these features we first turn to the eigenvalues of the neutrino mass matrix, then discuss the adiabatic approximation, and the adiabatic transfer of neutrino species. We return to the scalar field's dynamics in \cref{scalar-dynamics}.

\begin{figure}[!t]
\centering
    \includegraphics[keepaspectratio, width=0.75\textwidth]{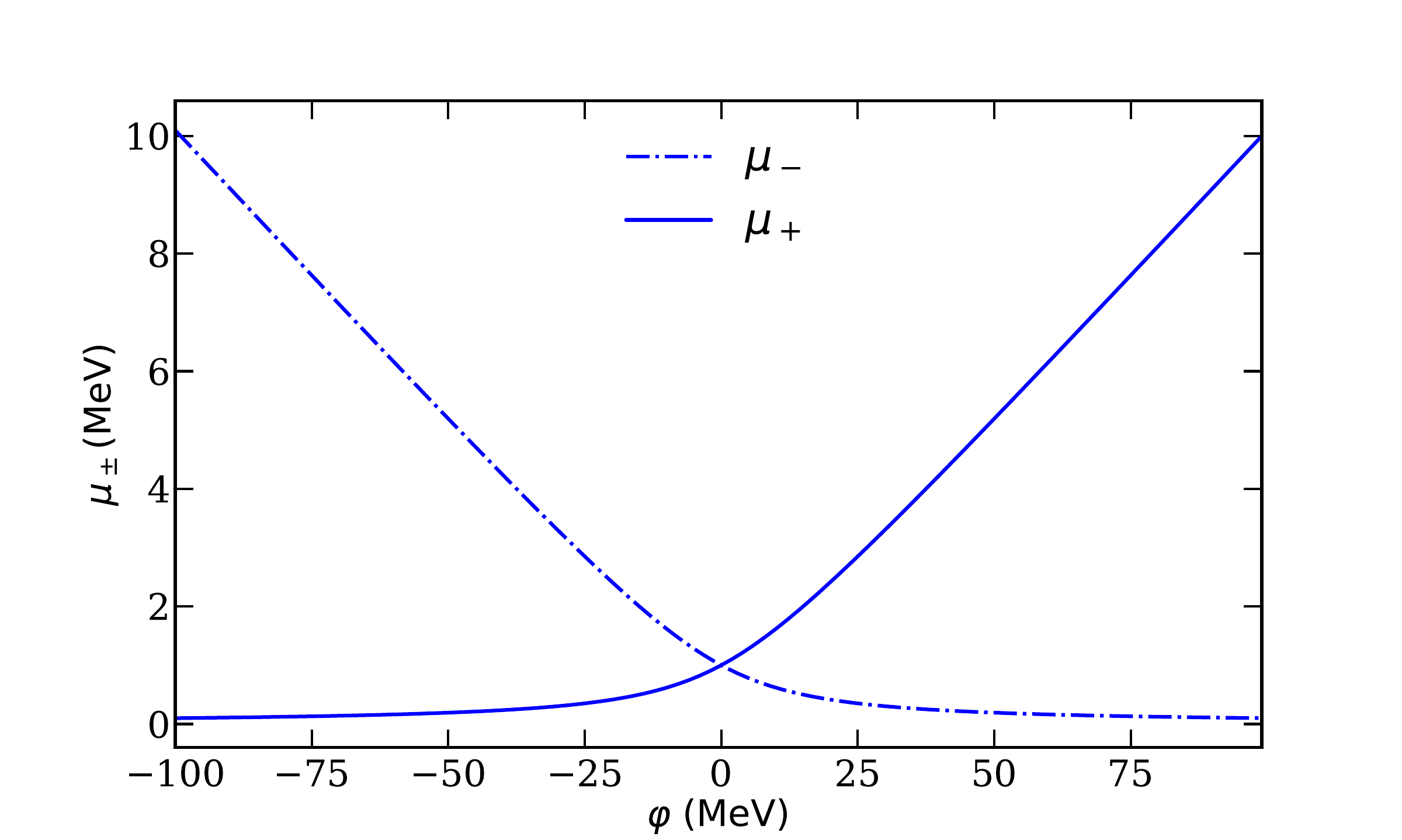}
\caption{The absolute value of \cref{eigenvalues} (i.e., neutrino masses) for $m_D = 1 ~ \rm MeV$ and $g =0.1$. At the zero-crossing point $\varphi = 0$ we see that both eigenvalues are equal to $m_D$ and the left- and right-handed neutrinos form a Dirac pair. For large positive $\varphi$, we see that $\mu_-$ approaches the seesaw limit, $m_D^2 / (g \varphi)$, while $\mu_+$ approaches $g \varphi$. The behavior is reversed for large negative $\varphi$. For large positive values of $\varphi$, $\ket{\nu_-}$ of \cref{mu-} acts as the `light neutrino' and $\ket{\nu_+}$ of \cref{mu+} as the `heavy neutrino', and vice versa for large negative values of $\varphi$.}
\label{fig:eigenvalue_plot}
\end{figure}

\subsection{Eigenvalues, masses, and states \label{sec:eigenvalues}} 
Let us consider the neutrino mass matrix at a fixed field value $\phi$, 
\begin{equation}
    \label{mass-matrix}
    M=\begin{pmatrix} 
        0 & m_D \\ 
        m_D & g \phi -m_N
    \end{pmatrix}
    =
    \begin{pmatrix} 
        0 & m_D \\ 
        m_D & g \phihat
    \end{pmatrix}~,
\end{equation}
with $m_D = y \vEW / \sqrt{2}$. In the second equality we have introduced $\phihat=\phi-\phi_c$ where $\phi_c=m_N/g$. 
The absolute values of the eigenvalues of \cref{mass-matrix} are
\begin{equation}
\label{eigenvalues}
    \mu_\pm = \frac{1}{2} \left( \sqrt{4 m_D^2 + (g \phihat)^2} \pm g \phihat \right)~,
\end{equation}
and the normalized eigenvectors in terms of the flavor basis $(\nu, n^c)$ are 
\begin{align}
\label{mu+}
    \ket{\nu_+}&= 
     \sin\theta \ket{\nu} +
     \cos\theta \ket{n^c}~,
    \\
\label{mu-}
   \ket{\nu_-}&= 
     \cos\theta \ket{\nu} -
     \sin\theta \ket{n^c}~,
\end{align}
where
\begin{equation}
\label{sin^2}
    \sin^2 \theta = \frac{\mu_-}{\sqrt{4m_D^2 + (g\phihat)^2}}~. 
\end{equation}
The coupling to  scalar-quanta in the mass basis is given by 
\begin{equation}
    \mathcal{L}_{\rm int} = (\sin^2\theta \nu_+\nu_+ + \cos^2\theta \nu_- \nu_- + 2 \sin\theta \cos\theta \nu_- \nu_+)\varphi   ~.
\end{equation}
There are both on- and off-diagonal couplings. 

Notice that for large and positive $\phihat$, the eigenvalues approach the approximate (see-saw) values 
\begin{equation}
    \label{positive-phi-masses}
    \begin{split}
    \mu_+ &\approx g \phihat~, \\
    \mu_- &\approx \frac{m_D^2}{g \phihat}~,
    \end{split}
\end{equation}
while (again for $\phihat \gg m_D/g$) the eigenvectors become 
\begin{align}
    \label{positive-phi-eigenvectors}
    \ket{\nu_+}&\approx 
     \ket{n^c}+ \tfrac{m_D}{g\varphi}  \ket{\nu}~,    
    \\
   \ket{\nu_-}&\approx 
     \ket{\nu}~ -
     \tfrac{m_D}{g\varphi} \ket{n^c}~.
\end{align}
We see that $\nu_+$ is the `heavy' mass eigenstate that is mostly composed of the sterile neutrino, while $\nu_-$ is the `light' mass eigenstate mostly composed of the active neutrino. The roles are reversed for large and negative $\phihat \ll -m_D/g$, where we find that 
\begin{equation}
    \label{negative-phi-masses}
    \begin{split}
    \mu_+ &\approx \frac{m_D^2}{g \phihat} ~,\\
    \mu_- &\approx g \phihat~,
    \end{split}
\end{equation}
with eigenvectors
\begin{equation}
    \label{negative-phi-eigenvectors}
    \begin{split}
    \ket{\nu_+}&\approx 
     \phantom{-}\ket{\nu}~+ \tfrac{m_D}{g\varphi}  \ket{n^c} ~,   
    \\
   \ket{\nu_-}&\approx 
     - \ket{n^c} +
     \tfrac{m_D}{g\varphi} \ket{\nu}~.
     \end{split}
\end{equation}
For $\phihat = 0$ the masses are degenerate, with $\nu_\pm$ forming a Dirac-pair with mass $|\mu_\pm| = m_D$. In \cref{fig:eigenvalue_plot} we plot the masses of the two eigenstates (absolute value of the eigenvalues) as a function of $\varphi$. 

When considering cosmology or particle kinematics (e.g., for a decay) it is convenient to track which neutrino species is heavy and which is light. We therefore introduce $\nu_L$ and $\nu_H$. For example, $\nu_L$ is defined by ({\it cf.} \cref{fig:eigenvalue_plot})
\begin{equation}
        \ket{ \nu_L} = \begin{cases}
            \ket{\nu_+} & \phi<\phi_c~\\
            \ket{\nu_-} & \phi> \phi_c  ~.
        \end{cases}
\end{equation}
It is similarly useful to introduce the mixing angle $\sin\theta_L$
\begin{equation}
    \sin\theta_L =  \braket{\nu_L}{n^c} = 
    \begin{cases}
           \cos\theta & \phi<\phi_c~\\
            \sin\theta & \phi> \phi_c  ~,
    \end{cases}
\end{equation}
such that $\sin^2\theta_L\leq 1/2$. 
\subsection{Adiabatic approximation}
Let us next consider the equations of motion for the fields $\nu$ and $n^c$. It is convenient to assemble them into a vector  $N= (\nu, n^c)^T$. Then the equations of motion read, 
\begin{equation}   
    \label{eom}
    \qty[\iu \partial \cdot \overline{\sigma} - \underline{M}(t)] N=0~.
\end{equation}
All underlined matrices act in flavor space (as opposed to $\overline{\sigma}^\mu$ which acts on spinor indices). These equations may be conveniently re-cast in the instantaneous eigenbasis by diagonalizing $\underline{M}(t)$ at each instant in time. Let us introduce the rotation matrix
\begin{equation}
    \underline{R}=\begin{pmatrix}
        \phantom{-}\cos\theta(t) & \sin\theta(t) \\
        -\sin\theta(t) & \cos\theta(t) 
      \end{pmatrix}~.
\end{equation}
Inserting $\underline{R}^T(t)\underline{R}(t)$ into \cref{eom}, and acting from the left with $\underline{R}(t)$ we obtain
\begin{equation}   
    \label{eom-diag-1}
    \qty[\iu \partial \cdot \overline{\sigma} ~\underline{1} -\underline{\cal M}(t) + \iu \underline{R}(t)\partial_t \underline{R}^T(t)] {\cal N}=0~,
\end{equation}
where ${\cal N}=(\nu_+, \nu_-)^T$ are the instantaneous mass eigenstates and $\underline{\cal M}(t)={\rm diag}[\mu_+(t), \mu_-(t) ]$. With some elementary trigonometric identities, one can show that $i \underline{R}(t)\partial_t 
\underline{R}^T(t)= (\partial_t \theta) \sigma_2$ with $\sigma_2 = \left(\begin{smallmatrix} 0 & -\iu \\ \iu & 0 \end{smallmatrix}\right)$. We then obtain
\begin{equation}   
    \label{eom-diag-2}
    \qty[\iu \partial \cdot \overline{\sigma}~ \underline{1} - \underline{\cal M}(t) + \dot{\theta}(t) \underline{\sigma_2} ] {\cal N}=0~.
\end{equation}
In the limit of slowly varying $\theta(t)$, we may drop the third term in \cref{eom-diag-2}. This is the adiabatic approximation and gives, 
\begin{equation}   
    \label{eom-diag-3}
    \qty[\iu \partial \cdot \overline{\sigma}~ \underline{1} - \underline{\cal M}(t) ] {\cal N}=0~.
\end{equation}
The adiabatic approximation is valid when \cite{Zener:1932ws, Parke:1986jy}
\begin{equation}
    \label{adiabatic-inequality}
       m_D^2 \gg g \dot{\phihat} ~,
\end{equation}
which physically corresponds to the gap in the spectrum, $\Delta \sim m_D$, being larger than the rate of change of the Majorana mass $\dv{t} (m_N-g\phi)$. Whenever \cref{adiabatic-inequality} is satisfied jumps between the two instantaneous eigenstates are exponentially suppressed. 

As an aside, since the Majorana mass of $n^c$ crosses zero, it is interesting to ask if non-perturbative particle production can take place; we find that it does not. This issue has been extensively studied in the context of inflation \cite{Abbott:1982hn, Greene:1998nh, Greene:2000ew, Adshead:2015kza} and for a single fermion particle production generically occurs whenever a particle's mass vanishes. This is because around this limit, the gap between positive and negative energy solutions (to the Klein-Gordan, Dirac, or other relativistic wave equation) vanishes. In the present context the off-diagonal Dirac mass supplies a non-zero gap even for a vanishing Majorana mass. It is easily checked that whenever \cref{adiabatic-inequality} is satisfied non-perturbative particle production is exponentially suppressed. We therefore conclude that for the dynamics of interest the adiabatic approximation is valid, there is no particle production, and the $n_+$ and $n_-$ populations are conserved. This conservation is only violated by heavy neutrino decays $\nu_H \rightarrow \nu_L + \phi$. 

%
\subsection{Adiabatic transfer \label{adiabatic-transfer}}
In the adiabatic approximation the populations of each instantaneous eigenstate (i.e., the labels $\pm$) are separately conserved. Therefore, if $\phi(t)$ crosses $\phi_c$ (or equivalently if $\phihat$ crosses $0$), then light-neutrinos will become heavy. This gives a non-thermal mechanism for the generation of heavy neutrinos from a thermal or relic population of light neutrinos. The heavy neutrino can then decay into the light neutrino and scalar, producing a population of relativistic light neutrinos
\begin{equation}
   \nu_H \to \nu_L + \phi~,
\end{equation}
where $\nu_H$ is the heavier instantaneous eigenstate at a given time. 

The decay rate in the rest frame for $\nu_H \rightarrow \nu_L +\phi$ is given by 
\begin{equation}
    \Gamma = \frac{g^2 \cos^2\theta \sin^2\theta}{16\pi}\qty(\frac{m_H^4-m_L^4}{m_H^3})~. 
\end{equation}
where 
\begin{equation}
    \sin^2\theta\cos^2\theta = \frac{m_D^2}{(g\varphi)^2 + 4 m_D^2}~. 
\end{equation}
The mass of the heavy and light states are also functions of the scalar field. In the limit where $g \varphi \ll 2m_D$ we find, 
\begin{equation}
    \Gamma \simeq \frac{g^4}{8\pi}|\varphi|\qty(1 + O\qty(\frac{g\varphi}{m_D}))~.
\end{equation}
Decays are almost instantaneous on cosmological timescales. This results in most decays occurring near the mass degenerate limit, and very little energy being released. 

\section{Scalar field dynamics \label{scalar-dynamics}}
Having established how the fermion masses change with $\phi$, we now turn to the equations of motion for $\phi$
\begin{equation}
    \label{phi-eom}
    \ddot{\phi} + 3 H \dot{\phi} + V'(\phi, t) = 0~.
\end{equation}
Here, $V(\phi,t)$ contains two contributions, 
\begin{equation}
    V(\phi,t) = V_0(\phi) + V_{\nu}(\phi,T)~,
\end{equation}
where $V_{\nu}(\phi,T)$ is the energy density of neutrinos whose mass depends on $\phi$. 

The form of $V_{\nu}(\phi,T)$ depends on whether or not neutrinos are in thermal equilibrium with the Standard Model bath. Above the temperature of neutrino decoupling $T\geq T_{\nu, {\rm dec}} \approx 1~{\rm MeV}$, light mostly-active neutrinos are in thermal equilibrium and dynamically adjust their phase space distribution in response to changes to their mass.\!\footnote{As we discuss in \cref{sec:higher-temps}, even above $T_{\nu,{\rm dec}}$ neutrino interactions rates are sometimes too slow to maintain thermal equilibrium.} Below $T_{\nu,{\rm dec}}$   neutrinos are a relic species, and are essentially inert. In this limit the neutrino number density is conserved. We now discuss the relic and thermal potentials separately.

\subsection{The relic potential}
After neutrino decoupling, the momentum distribution of neutrinos is fixed, and (assuming a standard cosmology for neutrino decoupling) is described by a massless Fermi-Dirac distribution with $T \simeq T_{\nu, \rm dec}$. As the universe expands, the momentum distribution redshifts and the number distribution scales as (see \cite{Fardon:2003eh,Kaplan:2004dq,Ghalsasi:2014mja} for a discussion in the context of mass-varying neutrinos)
\begin{equation}
    \label{fermi-dirac}
    \begin{split}
        n(k, T) &= n_F \left(k \frac{T_{\nu,\rm dec}}{T}, T_{\nu,\rm dec} \right) = \frac{1}{\e^{k / T} + 1} ~, 
    \end{split}
\end{equation}
with an integrated number density that scales as $(T/T_{\nu,\rm dec})^3$ as the universe cools. 
This phase space distribution then contributes to a ``relic potential''. Changing the mass of all neutrinos shifts the energy density by 
\begin{equation}
    \mathcal{E}(\phi) - \mathcal{E}(\phi = 0) = \frac{1}{2} m_\phi^2 \phi^2 + \int \frac{d^3 k}{(2 \pi)^3} n_\pm(k) \left(\sqrt{k^2 + \mu_\pm^2} - \sqrt{k^2 + \mu_\pm^2(\phi=0)}\right)~.
\end{equation}
If we assume an arbitrary number density of $+$ and $-$ states then the relic potential is given by 
\begin{equation}
    V_\text{relic}(T) = \int \frac{d^3 k}{(2 \pi)^3} n_{\mu_-}(k, T) \sqrt{k^2 + \mu_-^2}  + \int \frac{d^3 k}{(2 \pi)^3}n_{\mu_+}(k, T) \sqrt{k^2 + \mu_+^2} ~.
\end{equation}
In the limit of a single relic species $(\pm)$ with a number density given by \cref{fermi-dirac}, the resulting effective potential simplifies in the limit of $T\gg \mu_{\pm}$ and $T\ll \mu_{\pm}$. These two limiting forms are given by 
\begin{equation}
\label{V_relic}
    V_{\rm relic} \simeq 
    \begin{cases}
        \frac{3 \zeta(3)}{4\pi^2} T^3 \mu_\pm(\phi) & T\ll \mu_{\pm} \\[8pt]
        \frac{1}{48} T^2 \mu_\pm^2(\phi) & T\gg \mu_{\pm}~.
    \end{cases}
\end{equation}

\subsection{Abundance of light scalars}
The scalars we consider are light, and could be produced thermally via $\bar{\nu}_L \nu_L \rightarrow \phi \phi$ in the early universe. The thermally averaged cross section (assuming $m_L \ll T$) is dominated by heavy neutrino exchange (see \cref{sec:SN-cooling}) and is given parametrically by
\begin{equation}
    \Gamma =\langle \sigma n v \rangle \sim \frac{T^3}{m_H^2} \qty(\frac{m_L}{m_H})^2 ~.
\end{equation}
Setting this rate equal to Hubble in radiation domination we find the decoupling temperature $T_*$,  
\begin{equation}
    \frac{T^2_*}{M_{\rm Pl}} \sim \frac{T^3_*}{m_H^2} \qty(\frac{m_L}{m_H})^2  \implies  T_* \sim \frac{m_H^4}{M_{\rm Pl} m_L^2} ~.
\end{equation}
Taking $m_L \leq m_\nu \sim 0.1~{\rm eV}$ this gives, 
\begin{equation}
    T_* \gtrsim 40 ~{\rm GeV} \times \qty(\frac{m_H}{1~{\rm GeV}})^4~. 
\end{equation}
At this epoch $m_H$ will be dominated by $|g\varphi_{0,\rm est}|$. As we will see in what follows, (specifically see \cref{phi_0}) this then guarantees that at $T_*$ the relic density of $\phi$ particles that have frozen out will be negligible at BBN. We therefore do not consider the thermal population of scalars in the rest of our analysis.

\section{Cosmological scenarios \label{sec:cosmo-const}}
The presence of background potentials that are proportional to the neutrino density can dramatically alter the scalar field dynamics. The precise cosmological history depends qualitatively on the the initial conditions, and the ordering of various time scales. Specifically we will be interested in the value of the field's initial condition, $\phi_0$, relative to the critical field value $\phi_c$. In this section we describe how the cosmological history is modified. 

The mass of the scalar field determines at what temperature oscillations onset. 
This always occurs during radiation domination such that $H=\gamma T^2/M_{\rm pl}$ with $\gamma= \sqrt{\pi^2 g_*(T)/90}$. Solving for $3H =m_\phi$ then gives,  
\begin{equation}
     T_{\rm osc} = \sqrt{\frac{m_\phi M_{\rm pl}}{3\gamma(T_{\rm osc})}} \approx  4 \times 10^5 {\rm ~ eV} \bigg(\frac{0.65}{\gamma(T_{\rm osc}) }\bigg)^{1/2} \bigg(\frac{m_\phi}{10^{-16}~{\rm eV}}\bigg)^{1/2} ~,
 \end{equation}
where we have benchmarked the number of effective degrees of freedom using $g_*(T_{\rm osc})=3.91$. For $m_\phi \lesssim 10^{-14}~{\rm eV}$, we have $T_{\rm osc} \lesssim T_{\nu,\rm dec}$ and can therefore use the relic potential for the entirety of the scalar field's dynamics. We discuss the case of $m_\phi \gtrsim 10^{-14}~{\rm eV}$ in \cref{sec:higher-temps}. For $m_\phi\lesssim 10^{-14}~{\rm eV}$ the analysis of the next two sections applies immediately. For heavier scalars, our constraints on parameter space may be viewed as conservative, since any acceptable cosmology will eventually lead to a relic neutrino population at $T<T_{\rm \nu,dec}$. We discuss the dynamics of the scalar field during epochs where neutrinos are in thermal equilibrium qualitatively in \cref{sec:higher-temps}.

\subsection{Small amplitude initial conditions }
For large-$m_N$ or small-$g$ the value of $\phi_c$ can be so large that a standard misalignment mechanism is valid from $T_{\rm osc}$ until today. This scenario is simple to analyze, and its domain of validity can be easily inferred by blue-shifting the present-day dark matter density to $T_{\rm osc}$. 

By assumption (in order to have ``small" initial conditions) we require $|\phi_0|\leq \phi_c$. The potential is well approximated as quadratic, and the amplitude will red-shift in time like $A_\phi\sim (T/T_{\rm osc})^{3/2}$. The relic density today is given roughly by 
\begin{equation}
\label{phi_0-condition}
    \rho_{\rm DM,now} = \frac{1}{2} m_\phi^2 \phi_0^2 \qty(\frac{T_{\rm now}}{T_{\rm osc}})^3 
    ~.
\end{equation}
Demanding that $\phi_0$ in \cref{phi_0-condition} is less than $\phi_c$ gives an upper-bound on the relic density as a function of $m_\phi$, $m_N$, and $g$. Therefore, using $\rho_{\rm DM,now}= 9 \times 10^{-12}~{\rm eV}^4$,  in order for $|\phi_0|\leq \phi_c$ can be written as
\begin{equation}
    \label{boring-inequality}
    m_N \gtrsim  3 \times 10^{14}~{\rm GeV} \qty(\frac{10^{-16}~{\rm eV}}{m_\phi})^{1/4} \qty(\frac{g}{0.1})~. 
\end{equation}
This limit leads to very heavy right-handed neutrinos, which will therefore be Boltzmann suppressed and out of the bath well before BBN. Notice, also, that the Majorana mass of the right-handed neutrino is so large that even for $g\sim O(1)$ there will be no observable DiNO signatures \cite{Krnjaic:2017zlz,Brdar:2017kbt}.

\subsection{Large negative amplitude initial conditions \label{sec:large-negative}}
For lighter right-handed neutrinos the inequality \cref{boring-inequality} will be violated and the Majorana neutrino mass will naively cross zero. In reality, at early epochs the neutrino density is so large that the relic potential can cause the $\phi=\phi_c$ region to become kinematically forbidden. During the epoch following neutrino decoupling, our potential is  given by
\begin{equation}
    V(\phi, t) = V_0(\phi) + V_{\rm relic}(\phi, T)~,
\end{equation}
with $V_0(\phi)$ given by \cref{V_0} and $V_{\rm relic}$ given by \cref{V_relic}, with $\mu = \mu_+$ since according to \cref{negative-phi-masses} this is the light eigenstate for large negative values of $\phi$. 
 We take the heavy eigenstate to be Boltzmann-suppressed, assuming that at early times (when the field is stuck by Hubble friction) that the Majorana mass, $|m_N-g \phi_0|$, is large compared to the temperature. 

\subsubsection*{Absence of Zero Crossings}
We begin by estimating initial conditions assuming (naively) that the energy density in the scalar field oscillations red-shifts like matter $\rho\sim T^3$. We will revisit this assumption below. Without loss of generality\footnote{If the field begins at $0<\phi_*<\phi_c$ then it will ``drop'' much earlier than $T\sim \sqrt{m_\phi M_{\rm pl}}$, and then get stuck by Hubble friction at its turning point at $\phi_0<0$.} we can assume the field begins at $\phi_0<0$. We then estimate the initial conditions using the density of dark matter today $\rho_{\rm DM,now}\approx 9 \times 10^{-12}~{\rm eV}^4$, and the relation, 
\begin{equation}
\label{phi_0}
    \phi_{0,\rm est} = \sqrt{\frac{2 \rho_{\rm DM,now}}{m_\phi^2}} \qty(\frac{T_{\rm osc}}{T_{\rm now}})^{3/2} \approx 3 \times 10^{24} ~ {\rm eV} \bigg(\frac{m_\phi}{10^{-16}~{\rm eV}}\bigg)^{1/4}, 
\end{equation}
where $T_{\rm now}= 2.3\times10^{-4}~{\rm eV}$. The initial field value is very large at $T_{\rm osc}$ as is the initial neutrino density. For such a large initial field value the potential is well-approximated by the bare potential, and we neglect the relic potential in the calculation of \cref{phi_0}. The left-turning point, $\phi_L$, is initially equal to $\phi_0$, but will decrease as the universe cools due to the red-shifting of the field's energy density. 

Zero crossings will occur if the right turning point crosses $\phi_c$. However, we find that the inequality $\phi_R < \phi_c$ is satisfied for all temperatures. In what follows we will assume that $m_D\gg T$ such that the low-temperature approximation to the relic potential is valid for $\phi$ close to $\phi_c$.\!\footnote{At sufficiently large $|\phi-\phi_c|$ one can always have $m_L\ll T$, such that the high temperature expansion of the relic potential must be used at large field values.} At early times, when $\phi_R$ is closest to $\phi_c$, the right-turning point is given by
\begin{equation} \label{no-zero-crossings}
    \phi_R \simeq \phi_c - \frac{2m_D}{g}\qty(\frac{m_D^2 n_\nu^2- \rho_{\rm DM}^2}{2m_D n_\nu \rho_{\rm DM}}) \simeq \phi_c - \frac{2 m_D}{g} \qty(\frac{m_D n_\nu }{m_\phi^2 \phi_L^2})~.
\end{equation}
In deriving \cref{no-zero-crossings}, we used the fact that $\phi_R$ is close to $\phi_c$ and that $\phi_L$ is very large and negative. Therefore, to compute $\phi_R$ we may set the energy density equal to the relic potential (neglecting the bare contribution), whereas for $\phi_L$ we set the energy density equal to the bare potential (neglecting the relic contribution).

\Cref{no-zero-crossings} implies that zero crossings do not occur with initial conditions to the left of $\phi_c$ for the highest temperatures. Notice that the term in parentheses is independent of $T$ for $n_\nu \sim T^3$ and $\phi_L^2 \sim T^3$. Therefore, in this high temperature regime, the right-turning point is also independent of temperature, and the field does not cross $\phi_c$ during its dynamical evolution. At lower temperatures, the field does not even approach $\phi_c$ and we conclude that zero crossings occur at no temperatures.

A possible exception to our analysis occurs when $m_D \ll T$, in which case the pre-factor of the relic potential scales as $T^2$ as opposed to the $\rho_{\rm DM} \sim T^3$ growth of the scalar field's energy density. This may then allow the energy density in the scalar field to overcome the potential barrier from the relic potential. We comment briefly on this possibility in \cref{sec:conclusions}. 

\subsubsection*{Modified Redshifting of the Dark Matter Energy Density}
Since the potential is highly anharmonic, the field will (generically) not red-shift like matter. Nevertheless, as we will now argue, over most of cosmic history the $\rho\sim T^3$ scaling is a very good approximation. There is a short epoch where the field loses energy density faster than standard dark matter, but this period of time is short and leads to an $O(1)$ change in the dark matter energy density, after which the field resumes its behavior scaling as $\rho \sim T^3$.  This $O(1)$ change can spoil concordance between estimates/measurements of the dark matter energy density at different cosmological epochs. 

The Hubble damping of a scalar field in a generic potential can be obtained by computing the time-averaged ratio of $\dot{\phi}^2$ relative to $\rho$ \cite{Turner:1983he}, 
\begin{equation}
\label{gamma-def}
    \gamma(\rho) = \frac{\langle \dot{\phi}^2 \rangle}{\rho} = 2 \frac{\int_{\chi_L}^{\chi_{\rm min}} (1-\frac{V}{V_{\rm max}})^{1/2} \dd \chi  + \int_{\chi_{\rm min}}^{\chi_{R}} (1-\frac{V}{V_{\rm max}})^{1/2} \dd \chi ~}{\int_{\chi_L}^{\chi_{\rm min}} (1-\frac{V}{V_{\rm max}})^{-1/2} \dd \chi  + \int_{\chi_{\rm min}}^{\chi_{R}} (1-\frac{V}{V_{\rm max}})^{-1/2} \dd \chi}~, 
\end{equation}
where we re-scale the field $\psi = g\phi/(2m_D)$, and introduce
\begin{equation}
    \chi = \psi -\psi_c = \frac{g}{2 m_D}\qty(\phi-\phi_c)~. 
\end{equation}
The turning points, $\chi_{L,R}$ are a function of $\rho$. Assuming rapid oscillations (relative to Hubble) the time-averaged energy density obeys the non-linear differential equation 
\begin{equation}
    \label{eq:Turner}
    \dv{\rho}{t} = -3 H \gamma(\rho) \rho ~, 
\end{equation}
which can be solved given a solution $\gamma(\rho)$. To determine how $\chi_{\rm min}$ and the turning points $\chi_{L, R}$ change over time, we will investigate the evolution of the shape of the potential and the oscillations of the scalar field with temperature. The shape of the potential is characterized by two dimensionless ratios
\begin{equation}
\label{kappa}
    \kappa = \frac{g^2 n}{4 m_\phi^2 m_D} ~,
\end{equation}
\begin{equation}
\label{psi_c}
    \psi_c = \frac{g\phi_c}{2 m_D}~. 
\end{equation}
In terms of these variables, the potential is given by
\begin{equation}
\label{chi-potential-left}
    V(\chi) = 
    \begin{cases}
        \frac{4 m_D^2 m_\phi^2}{g^2} \left(\frac{1}{2} (\chi + \psi_c)^2 + \kappa \lambda_+ \right) & T\ll m_D \lambda_+ \\[8pt]
        \frac{4 m_D^2 m_\phi^2}{g^2} \left(\frac{1}{2} (\chi + \psi_c)^2 + \frac{m_D T^2}{48 n} 
        \kappa \lambda_+^2 \right) & T\gg m_D \lambda_+~~,
    \end{cases}
\end{equation}
where
\begin{equation}
\label{lambda}
    \lambda_\pm = \chi \pm \sqrt{\chi^2 + 1}~.
\end{equation}

Notice that $\kappa \propto n \propto T^3$ is largest at high temperatures. When $\kappa$ is large the minimum of the potential is pushed to large negative field values, which leads to a light neutrino mass $m_L \ll m_D$. At very high temperatures, where the relativistic approximation in \cref{V_relic} and \cref{chi-potential-left} applies, the minimum scales as $\chi_{\rm min} \sim (\kappa m_D/T)^{1/4} \sim T^{1/2}$. As the temperature decreases, $\chi_{\rm min}$ shifts towards the origin, and $m_L$ increases until $m_L(\chi_{\rm min}) \gg T$. At this point the non-relativistic approximation to \cref{V_relic} and \cref{chi-potential-left} is appropriate, and the potential is globally well approximated by 
\begin{equation}
\label{potential-approximation}
    V(\chi) \simeq \frac{1}{2} (\chi+\psi_c)^2 - \frac{\kappa}{2 \chi}  \qq{for} \kappa, \psi_c \gg 1~,
\end{equation}
whose minimum occurs at\footnote{This result is derived by taking \cref{potential-approximation}, counting $\chi_{\rm min} \sim \kappa^{1/3} \sim \psi_c \sim \lambda$, and expanding in the $\lambda\rightarrow \infty$ limit.}
\begin{equation}
\label{chi-min-left}
    \begin{split}
    \chi_{\rm min} &\simeq \frac{1}{3} \Bigg(\frac1{2^{2/3}}(\sqrt{27\kappa  \left(27 \kappa +8 \psi_c^3\right)}-27 \kappa -4\psi_c^3)^{1/3}\\
    &\hspace{0.2\linewidth}+\frac{2^{2/3}\psi_c^2}{(\sqrt{27\kappa  \left(27 \kappa +8 \psi_c^3\right)}-27 \kappa -4\psi_c^3)^{1/3}}- \psi_c\bigg)~.
    \end{split}
\end{equation}
One can clearly see by inspection that there are three different qualitative regimes: $\kappa^{1/3}\gg \psi_c$, $\kappa^{1/3} \sim \psi_c$, and $\kappa^{1/3} \ll \psi_c$. For example, when $\kappa^{1/3} \gg \psi_c$ the minimum scales $\chi_{\rm min} \simeq -(\kappa/2)^{1/3}$, when $\kappa^{1/3} \sim \psi_c$ \cref{chi-min-left} must be used, and when $\kappa^{1/3}\ll \psi_c$  then $\chi_{\rm min}\simeq -\psi_c$.   

\begin{figure}[b]
    \centering
    \begin{subfigure}[t]{0.32\textwidth} 
        \centering
        \includegraphics[keepaspectratio, width=\textwidth]{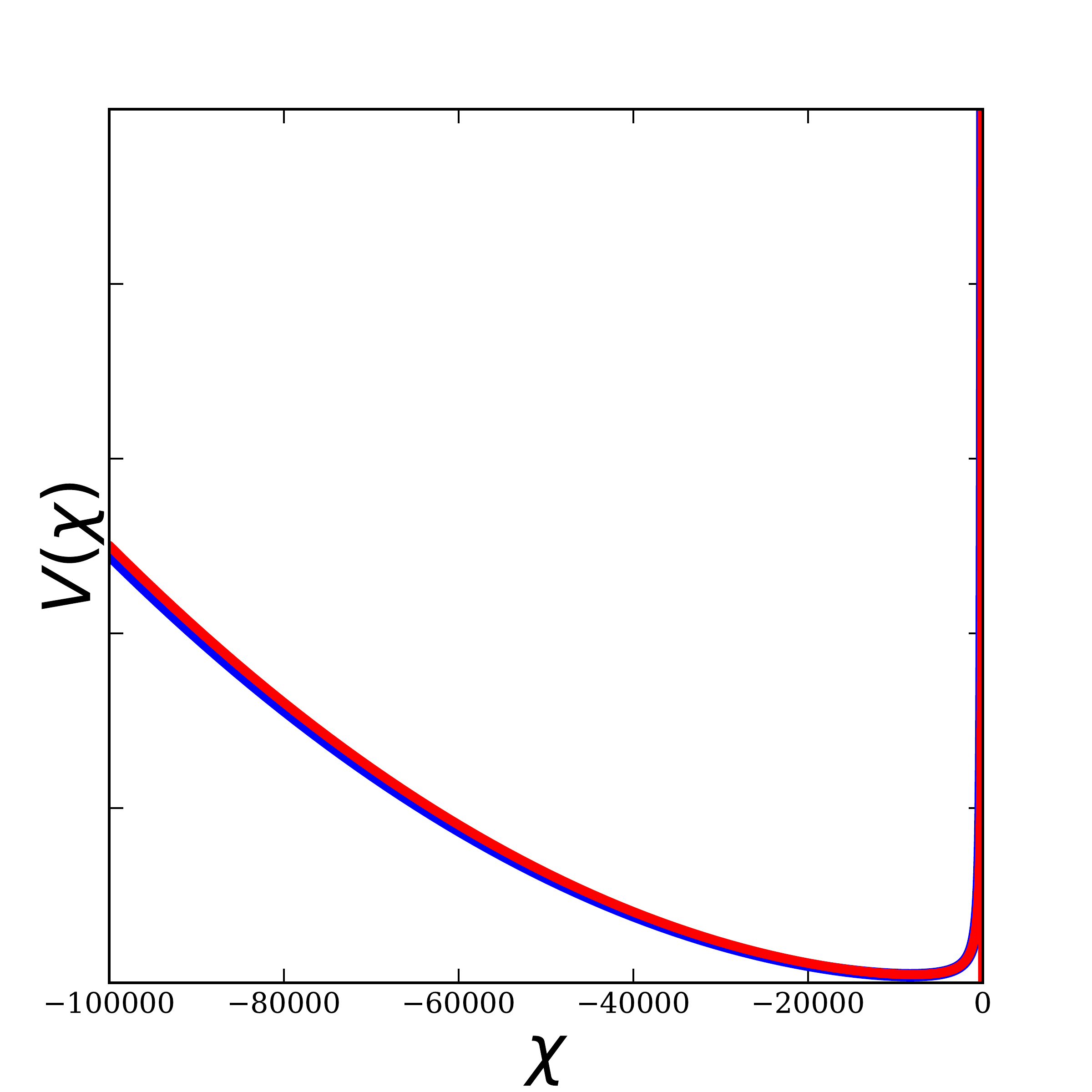}
        \caption{$A_L \gg \kappa^{1/3} \gg \psi_c$}
        \label{fig:imga}
    \end{subfigure}
    \begin{subfigure}[t]{0.32\textwidth} 
        \centering
        \includegraphics[keepaspectratio, width=\textwidth]{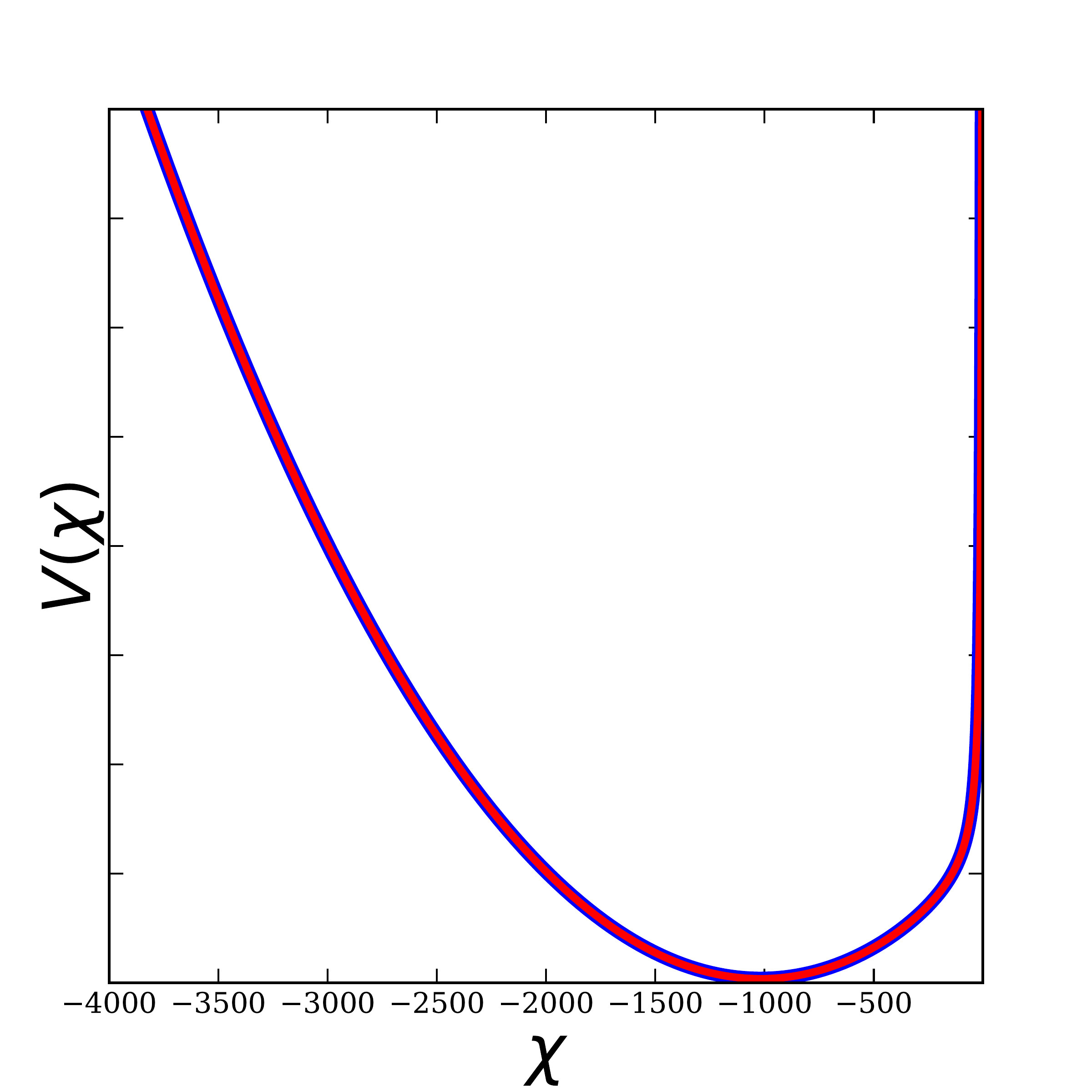}
         \caption{$A_L \gtrsim \psi_c \gtrsim \kappa^{1/3} $}
         \label{fig:imgb}   
    \end{subfigure}
    \begin{subfigure}[t]{0.32\textwidth} 
        \centering
         \includegraphics[keepaspectratio, width=\textwidth]{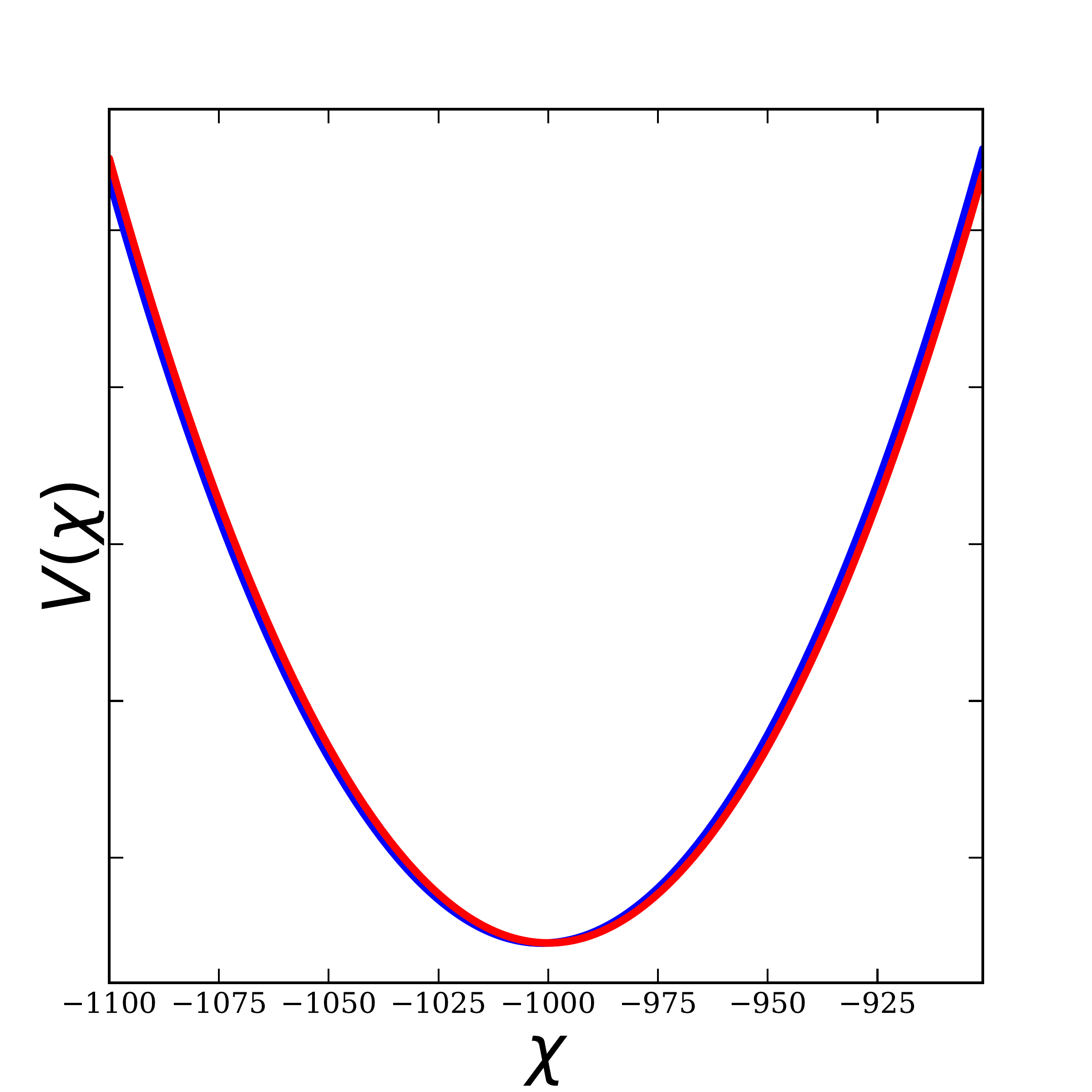}
         \caption{$A_L\ll \psi_c \lesssim \kappa^{1/3} $}
         \label{fig:imgc}   
    \end{subfigure}
\caption{Different shapes of the (dimensionless) potential throughout cosmic history for $\chi_0 < 0$ and $\psi_c=10^3$. Note the different scales on the abscissa ($x$-axis) between \textbf{(a)}-\textbf{(c)}, which are representative of $A_L$. The full potential is plotted in blue, and the approximation from \cref{potential-approximation} is plotted in red. \textbf{(a)} Initial conditions are such that the left turning-point is large $\chi_L \gg \chi_{\rm min}$. The scalar field is reflected at the origin and spends almost all of its time in a harmonic potential. \textbf{(b)} Hubble friction reduces the amplitude of oscillations until it is comparable to $\chi_{\rm min}$. At this stage the steep ``brick wall'' barrier causes the field to have a larger average kinetic energy than it otherwise would in a symmetric harmonic potential. \textbf{(c)} Eventually the amplitude becomes sufficiently small that the right turning-point is far removed from the ``brick wall'' and the field oscillates in a symmetric harmonic potential. }
\label{fig:shapes_of_potentials}
\end{figure}

 These different scalings lead to different shapes of the potential.
 In \cref{fig:shapes_of_potentials} we show the different shapes that one encounters as the universe cools. The plots are shown at different ``magnifications'', which are set by the amplitude of oscillations, %
\begin{equation}
     A_L=|\chi_L-\chi_{\rm min}|~.
\end{equation}

As shown in \cref{fig:imga}, at early times $A_L \gg \kappa^{1/3} \gg \psi_c$ and the potential looks like a harmonic oscillator with a steep barrier near $\chi\sim O(1) \ll A_L$. Since $|\chi_{\rm min}| \ll A_L$, the barrier and the minimum occur at roughly the same location. As the universe cools, $A_L$ drops like $T^{3/2}$ (with some corrections which we discuss below), and eventually $A_L$ becomes comparable to $\chi_{\rm min}$ when $A_L \gtrsim \psi_c \gtrsim \kappa^{1/3}$. During this epoch the shape of the potential resembles a harmonic oscillator centered near $\chi_{\rm min} \simeq -\psi_c$. The steep barrier near $\chi\sim O(1)$ is well separated (in units of $A_L$) from the minimum.  Finally, Hubble friction causes the amplitude to become sufficiently small that the steep barrier is energetically inaccessible (i.e., very far away in units of $A_L$), and the dynamics are well approximated by the bare potential.

We now return to the red-shifting of the dark matter energy density and  \cref{gamma-def}. In the early stages of the universe (\cref{fig:imga}), the amplitude of the field is very large compared to the minimum of the potential $\chi_L \gg \chi_R \simeq \chi_{\rm min}$. As a result, the first integral (i.e., the ``left'' portion of the oscillation) in both the numerator and denominator of \cref{gamma-def} provide the dominant contributions. The minimum of the potential is very close to $\chi=0$ (i.e., $\chi_{\rm min}/\chi_L \ll 1$), and we can think of the potential as a harmonic oscillator with a ``brick wall'' at $\chi=0$. Since the reflected dynamics on the half-interval $[\chi_L,0]$ are equivalent to the dynamics that would occur in a symmetric oscillator centered at $\chi=0$, we find $\gamma = 1 + O(\chi_{\rm min}/\chi_L)$ and the field red-shifts like matter.

As shown in \cref{fig:imgb}, the left-turning point $\chi_L$ decreases due to Hubble friction as the universe cools. In this epoch $A_L \sim \psi_c \sim \kappa^{1/3}$ and the left-amplitude, $A_L= |\chi_L-\chi_{\rm min}|$, becomes comparable to the right-amplitude, $A_R=|\chi_R-\chi_{\rm min}|$.
Both the left- and right-oscillations provide $O(1)$ contributions to the integrals in \cref{gamma-def}. In this regime, the ``brick wall'' is now offset from the minimum of the potential which has shifted to $\chi_{\rm min} \simeq -\psi_c$. Now, when the fields bounces off the wall at $\chi = 0$, it does so in the right-portion of its oscillation, which increases the average kinetic energy ($\langle \dot{\phi}^2 \rangle$) per oscillation cycle.  This leads to a short epoch where $\gamma \gtrsim 1$ ({\it cf.}~\cref{gamma-BW}).  Consequently, there is a sudden decrease in the scalar field's energy density relative to the naive expectation of $\rho \propto (T/T_{\rm osc})^3$ during this epoch. 

Eventually, as shown in \cref{fig:imgc}, this epoch ends when $A_R\sim A_L \ll \psi_c \sim \chi_{\rm min} $. At this point the steep barrier from the relic potential becomes energetically inaccessible, and the field oscillates in its symmetric bare potential about the minimum.  In this epoch $\gamma=1$, and the scalar field's energy density again red-shifts like matter. 

To compute the drop in the scalar field's energy density (relative to the typical scaling with matter) during the epoch depicted in \cref{fig:imgb} we solve \cref{eq:Turner}. It is convenient to model the relic potential as a ``brick wall'' as alluded to above. This amounts to a reflective boundary condition at $\chi=0$ in addition to the bare scalar potential. 
Within this approximation, \cref{gamma-def} may be re-written as (using $y=(\chi-\chi_{\rm min})/A_L$)
\begin{equation}
    \begin{split}
    \label{gamma-BW}
    \gamma_{\rm BW}(c) &\equiv 2\frac{ \int_0^{1} \qty(1-y^2)^{1/2} \dd y  + \int_0^{c}\qty(1-y^2)^{1/2} \dd y }{\int_0^{1} \qty(1-y^2)^{-1/2} \dd y  + \int_0^{c}\qty(1- y^2)^{-1/2} \dd y }\\ 
    &=\frac{ \frac{\pi}{2}  + c \sqrt{1-c^2}-2\cot ^{-1}\left(\frac{c+1}{\sqrt{1-c^2}}\right)}{\frac{\pi}{2} + \sin^{-1}(c) }  ~, 
    \end{split}
\end{equation}
where $c\equiv \chi_{\rm min}/A_L$. The above equation applies for $0\leq c\leq 1$, whereas for $c\geq 1$, corresponding to the case where the wall is energetically inaccessible, we have $\gamma_{\rm BW}(c)=1$. The relationship between $c$ and the energy density $\rho$ is given by, 
\begin{equation}
\label{c-def}
    c=\frac{\chi_{\rm min}}{\chi_L} \simeq \frac{m_\phi \phi_c}{ \sqrt{2\rho}}= \frac{m_\phi m_N}{ g\sqrt{2\rho}}~.
\end{equation}
Notice that $\gamma(c=0) = 1$, which corresponds to the epoch shown in \cref{fig:imga} where the brick wall occurs at (or very close to) $\chi_{\rm min}$ .

To obtain the shift in the energy density away from the naive $\rho \sim T^3$ scaling expected for non-relativistic matter, we may compute 
\begin{equation}
    \Delta(\rho) = \int_{\rho}^{\rho_0}\qty[\frac{1}{\gamma(\rho')} -1] \dd \log(\rho') ~, 
\end{equation}
which leads to 
\begin{equation}
    \qty(\frac{\rho}{\rho_0})\e^{\Delta(\rho)} = \qty(\frac{T}{T_{\rm now}})^3~. 
\end{equation}
Using the brick wall model and \cref{c-def} to change variables from $\rho'$ to $c'$, we find that
\begin{equation}
    \label{brick-wall-delta}
    \Delta_{\rm BW}(c) = -2\int_{c_0}^{c}\qty[\frac{1}{\gamma_{\rm BW}(c')} -1] \dd \log(c') ~, 
\end{equation}
and integrating numerically we find $\Delta_{\rm BW}(c\geq 1) \approx 0.693$, which leads to a decrease of $\e^{-\Delta_{\rm BW}}\approx 0.5$. Therefore, the net result of the asymmetric potential generated by the neutrino background is a dark matter energy density which experiences a sudden drop by $\sim50\%$ relative to what would be obtained for a cosmic history where $\gamma=1$. 

The temperature at which this large drop in the dark matter energy density occurs can be estimated (using $A_L \sim |\chi_{\rm min}| \sim \psi_c$ as is appropriate for the epoch depicted in \cref{fig:imgb}) by equating $\phi_{0,\rm est} \qty(\frac{T}{T_{\rm osc}})^{3/2}=\phi_c$, which gives 
\begin{equation}
\label{T_drop}
    \frac{T_{\rm drop}}{T_{\rm now}} =  \qty( \frac{m_\phi \phi_c}{\sqrt{2 \rho_{\rm DM,now}}})^{2/3} = 0.4 ~ \times \qty(\frac{m_N}{1~{\rm GeV}})^{2/3} \qty(\frac{m_\phi}{10^{-16}~{\rm eV}})^{2/3} \qty(\frac{0.1}{g})^{2/3} ~. 
\end{equation}
For  $T_{\rm now} < T_{\rm drop} \lesssim T_{\rm CMB}$, i.e. when \cref{fig:imgc} corresponds to the current epoch, we expect that this will (badly) spoil concordance between CMB measurements and estimates of the dark matter relic abundance today.  While this parameter space is naively ruled out, our results motivate a detailed study of transient equations of state as we discuss in more detail in \cref{sec:conclusions}. 

By way of contrast, if $T_{\rm drop} < T_{\rm now}$, i.e. when \cref{fig:imga} corresponds to the current epoch, then the amplitude of oscillations are still large enough that the $O(1)$ effect has not occurred. However this then has consequences for present day measurements of neutrino mass which will vary rapidly as a function of time due to the large oscillations in this regime. Specifically, we expect $O(1)$ modifications to time-averaged neutrino oscillation parameters, such that existing constraints from DiNOs (sensitive to $\sim 5\%$ distortions \cite{Krnjaic:2017zlz,Brdar:2017kbt}) entirely rule out this region of parameter space.

\subsection{Large positive amplitude initial conditions \label{sec:large-positive}}
 Next, let us consider initial conditions in which $\phi_0 > \phi_c$. Unlike the case when $\phi_0< 0$, where the relic potential only served to modify the (effectively conservative) scalar field's dynamics, in this case the heavy- and light-neutrino eigenstates can exchange roles at some point in cosmic history. We will take $\phi_0 \gg \phi_c$ such that $\nu_H$ is Boltzmann suppressed, and our potential again has the form
\begin{equation}
    V(\phi, T) = V_0(\phi) + V_{\rm relic}(\phi, T)~,
\end{equation}
with $\mu = \mu_-$ since according to \cref{positive-phi-masses} this is the light eigenstate for large positive values of $\phi$. We will mostly be concerned with the epoch at CMB and those following it, so we take $T \ll m_D \lambda_-$ and our potential in terms of $\chi$ and $\psi_c$ has the form
\begin{equation}
\label{chi-potential-right}
    V(\chi) = \frac{4 m_D^2 m_\phi^2}{g^2} \left(\frac{1}{2} (\chi + \psi_c)^2 + \kappa \lambda_- \right)~,
\end{equation}
with $\lambda_-$ once again given by \cref{lambda}. 

The minimum of the potential is found by solving $V'=0$, or equivalently 
\begin{equation}
    \frac{\psi_c}{\kappa} + \qty(1-\frac{\chi_{\rm min}}{\sqrt{\chi^2_{\rm min}+1}}) = - \frac{\chi_{\rm min}}{\kappa} ~.  
\end{equation}
Clearly the behavior of the solution is controlled by the ratio 
\begin{equation}
    r= \frac{\psi_c}{\kappa}~.
\end{equation}
When $\kappa \gg 1/r^{3/2}$ (or equivalently $\kappa \ll \psi_c^{3}$), the location of the minimum is given by 
\begin{equation}
    \label{phi0_right:chi_min_small}
    \chi_{\rm min} \simeq \qty( \frac{1-r}{\sqrt{r(2-r)}})~.
\end{equation}
This formula is most useful when $r\sim O(1)$.

At high temperatures, $r\rightarrow 0$, and the minimum moves to large field values. It is convenient to expand $\lambda_- \simeq 1/(2\chi)$ in \cref{chi-potential-right}, and solve for the minimum. Doing so one finds, 
\begin{equation}
    \begin{split}  
    \label{phi0_right:chi_min_big}
    \chi_{\rm min} \simeq  \frac{1}{3} \Bigg[&~~\frac{ \kappa  R}{\sqrt[3]{\tfrac14\kappa ^2 \left(3 \left(\sqrt{81-24 R}+9\right)-4 R\right) R}}\\
    &\hspace{0.25\linewidth}-\sqrt[3]{\kappa  R}+\sqrt[3]{\tfrac14\kappa  \left(3 \left(\sqrt{81-24 R}+9\right)-4 R\right)}~~\Bigg]~,
    \end{split}
\end{equation}
where 
\begin{equation}
    R=r^3\kappa^2=\frac{\psi_c^3}{\kappa}~,
\end{equation}
and the expression in the square brackets is $\sim O(1/R^{1/6})$ as $R\rightarrow \infty$. The minimum therefore scales as $\kappa^{1/3}/R^{1/6}\sim \kappa^{1/6}$ at high temperatures. 

As we will see in what follows, the cosmology of a field trapped at $\chi\geq 0$ is much more dramatic than for a field whose initial conditions are taken to be $\chi<0$. For this reason, the red-shift dependence of the energy density is less interesting, and we instead focus on a different phenomenon in which the identity of neutrinos in the bath changes character.

\subsubsection*{Zero-crossings and overclosing the universe \label{sec:delayed-zero-crossings}}
As the universe cools $r$ increases, and is eventually $\sim O(1)$. Notice that for $r\sim O(1)$ the minimum of the potential, \cref{phi0_right:chi_min_small}, occurs close to $\chi=0$ and for $r\geq 1$ actually occurs at negative values of $\chi$. During this epoch the field will generically cross $\chi=0$. This has dramatic consequences for the subsequent cosmology. 

\begin{figure}
    \centering
    \begin{subfigure}[t]{0.475\textwidth} 
        \centering
        \includegraphics[keepaspectratio, width=\textwidth]{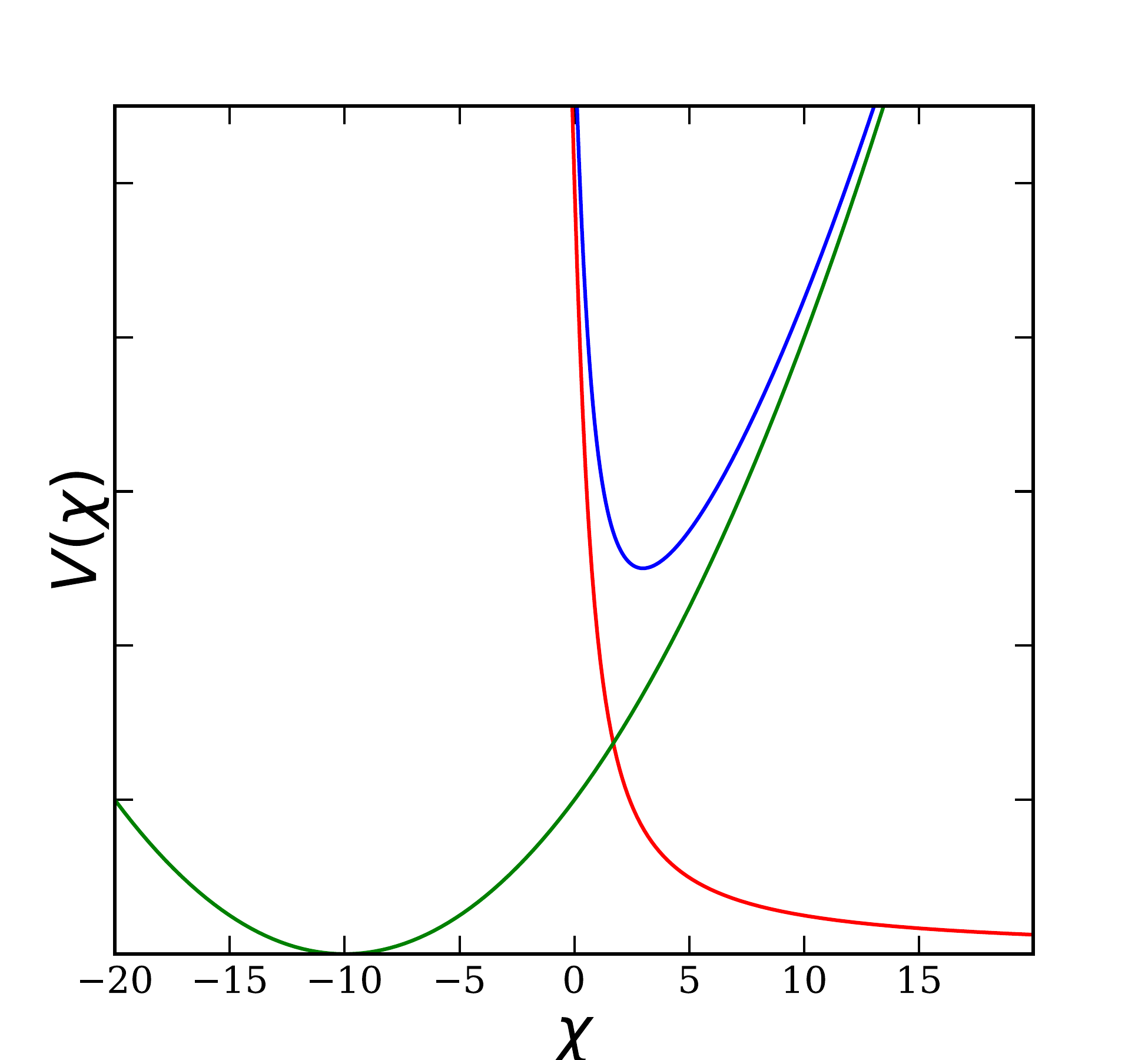}
        \caption{Before crossing zero ($T>T_{\rm cross}$).}
        \label{fig:pica}
    \end{subfigure}
    \begin{subfigure}[t]{0.475\textwidth} 
        \centering
        \includegraphics[keepaspectratio, width=\textwidth]{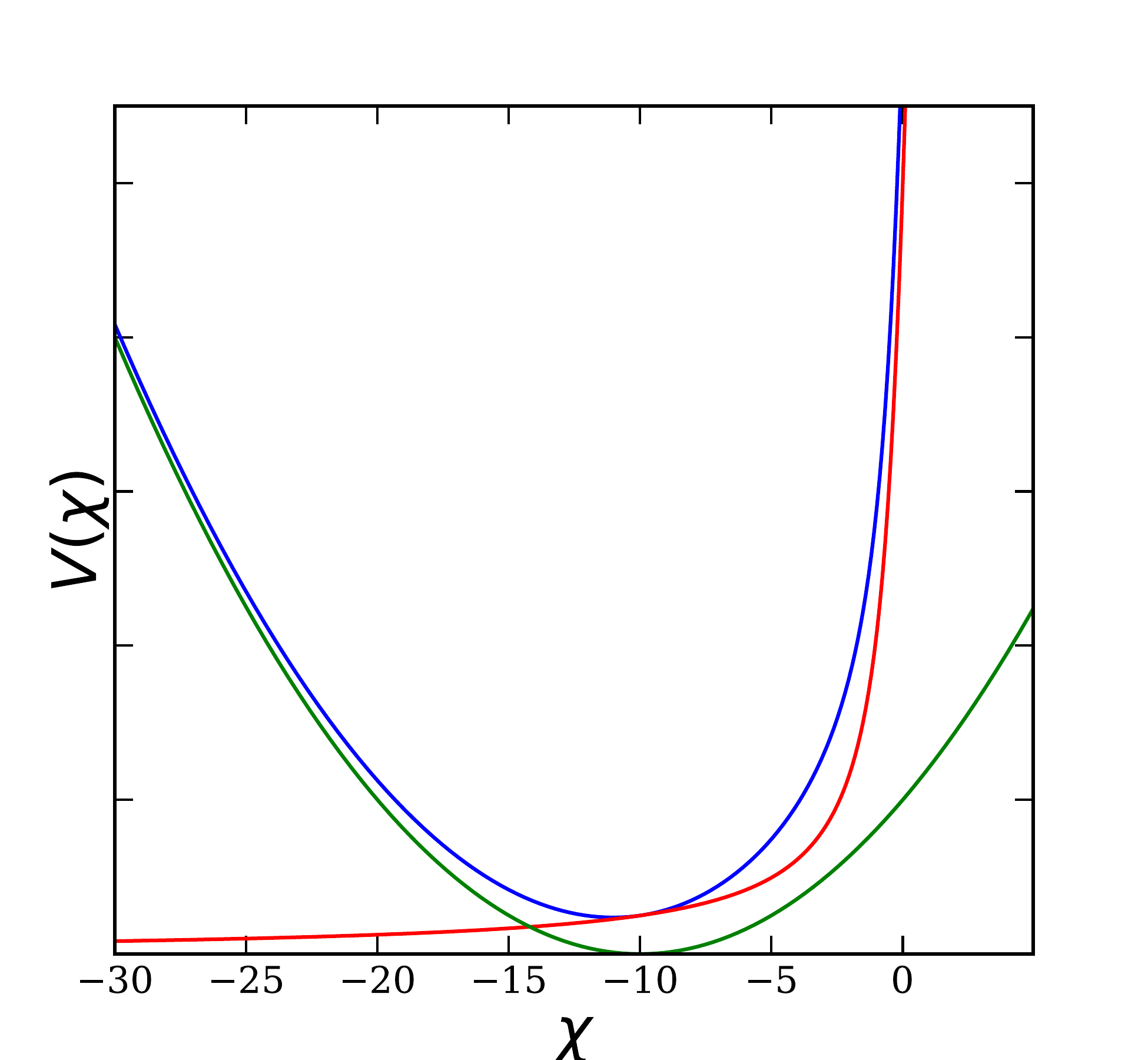}
         \caption{After crossing zero ($T<T_{\rm cross}$).}
         \label{fig:picb}   
    \end{subfigure} 
    \caption{Different shapes of the (dimensionless) potential throughout cosmic history for $\chi_0 > 0$. The full potential \cref{chi-potential-right} is plotted in blue, the relic potential is plotted in red, and the bare (harmonic) potential is plotted in green. \textbf{(a)} Before zero crossings, the relic potential serves as a steep ``wall'' that points in the opposite direction of the bare potential. This leads to small oscillations ``in the cup'' created by the two potentials. \textbf{(b)} After crossing zero, the relic potential ``flips'' and now drives $\chi$ towards large negative values. The net result is large oscillations centered around the minimum of the bare potential $\chi \sim -\psi_c$ (or $\phi \sim 0$).}
\end{figure}

As $\chi$ crosses 0, all of the neutrinos in the bath adiabatically convert from light-eigenstates to heavy-eigenstates. There is  then a short period of rapid decays (see \cref{adiabatic-transfer}). This can be summarized as, 
\begin{equation}
    \nu_{-,L} \underset{\rm adiabatic}{\longrightarrow}  \nu_{-,H} \underset{\rm decay}{\longrightarrow} ~\phi + \nu_{+,L}  ~,
\end{equation}
where we have tracked both ``heavy'' and ``light'', as well as the adiabatically conserved labels ``$-$'' and ``$+$''. This essentially instantaneously converts the $\nu_-$ population into a $\nu_+$ population, which effectively ``flips'' the relic potential such that it no longer opposes the bare potential and the minimum now occurs at $\phi\approx 0$ or equivalently $\chi\approx-\psi_c$ (see \cref{fig:picb}). The field is now misaligned from the new minimum with a large initial amplitude; oscillations begin immediately, and the energy density stored in the coherent oscillations of the field can overclose the universe.

Since the density of neutrinos is a monotonically decreasing function of temperature, so too is the $\chi_{\rm min}$ of \cref{phi0_right:chi_min_small}, and there will always exist a temperature $T_{\rm cross}$ such that zero-crossings occur. We numerically solve for the $m_N$ where $T_{\rm cross} = T_{\rm now}$ by determining when the left-turning point $\chi_L$ crosses zero, where $\chi_L$ satisfies the condition
\begin{equation}
\label{turning-point-condition}
    V(\chi_L) - V(\chi_{\rm min}) = \rho_{\rm DM, now}~,
\end{equation}
with $\chi_{\rm min}$ given by \cref{phi0_right:chi_min_small} and \cref{phi0_right:chi_min_big} in the appropriate regimes.\!\footnote{Where we use \cref{functional-approximation} to determine $m_D$ in terms of $m_N$, $m_\phi$, and g.} We find that over most of parameter space, the zero-crossing point is very well-approximated by simply taking the value of $m_N$ for which the minimum of the potential crosses zero, and dividing it by 200. The temperature for which the minimum crosses zero is easily determined by setting the $r=1$ ({\it cf.} \cref{phi0_right:chi_min_small}), which yields
\begin{equation}
\label{m_N-cross-now}
    m_{N_{\rm cross, now}} \approx \frac{1}{200} \frac{g^2 (3\zeta(3)/4\pi^2) T_{\rm now}^3}{2 m_\phi^2}~.
\end{equation}
However, for small enough values of $m_N$, the minimum of the potential becomes so close to zero that the zero-crossing point becomes essentially independent of $m_N$, and this approximation breaks down (see \cref{fig:parameter_space_right}).  We find that when $m_\phi\gtrsim 10^{-13}~{\rm eV}$ for $g = 10^{-4}$ or $m_\phi\gtrsim 10^{-10}~{\rm eV}$ for $g = 0.1$, the field will always cross zero before today. 

If $T_{\rm cross} > T_{\rm now}$, then zero crossings have certainly occurred in our cosmic history which implies that Majorana masses satisfying (see \cref{fig:parameter_space_right})
\begin{equation}
\label{delayed-zero-crossings}
    m_N \gtrsim  3 \times 10^{6} ~{\rm GeV}~ \qty(\frac{g}{0.1})^2 \qty(\frac{10^{-16}~{\rm eV}}{m_\phi})^2~,
\end{equation}
lead to cosmologies where the field passes through $\chi=0$ at some point in its evolution. 

After zero crossings occur, the potential has the shape shown in \cref{fig:picb}.  Near the minimum at $\phi = 0$, the potential is nearly zero, $V(\phi) \approx 0$, since the bare potential vanishes and neutrinos are nearly massless. The energy density stored in the field is therefore given by the value of the potential at zero-crossing, or 
\begin{equation}
\label{V_cross}
    \rho_{\rm DM, cross} = V_{\rm cross} = V_0(\phi_c) + V_{\rm relic}(\phi_c) = \frac{1}{2} m_\phi^2 \phi_c^2 + m_D n_\nu~.
\end{equation}
The oscillations are approximately harmonic, and we can assume a typical scaling of dark matter energy density with temperature $\rho_{\rm DM, now} \sim \left(\frac{T_{\rm now}}{T_{\rm cross}}\right)^3$. To avoid exceeding today's dark matter energy density, we therefore demand that 
\begin{equation}
\label{overclosure-condition}
    \qty[\frac{1}{2} m_\phi^2 \phi_c^2 + m_D n_\nu]\times  \left(\frac{T_{\rm now}}{T_{\rm cross}}\right)^3 \leq \rho_{\rm DM,now}~.
\end{equation}
Which leads to the condition 
\begin{equation}
\label{overclosure}
    m_N \lesssim 600~{\rm eV}~,
\end{equation}
where we use\footnote{Since the minimum is close to zero and the oscillations nearly harmonic after zero crossing, the `typical' seesaw relationship holds, so we replace the $m_D$ in \cref{V_cross} with $\sqrt{m_N m_\nu}$ in calculating this limit.} \cref{phi0_right:chi_min_small} to calculate $T_{\rm cross}^3$ by setting $r=1$ and replacing $m_N$ with $200 m_N$ based on the discussion surrounding \cref{m_N-cross-now}
\begin{equation}
    T_{\rm cross}^3 = \frac{400 m_\phi^2 m_N}{g^2 (3\zeta(3)/4\pi^2)} ~.
\end{equation}
Using the typical see-saw relationship $m_N\sim m_D^2/m_\nu$ one can see that $T_{\rm cross}^3 \sim \left(\tfrac{m_\phi^2}{m_D m_\nu}\right) m_D^3$ such that $T_{\rm cross} \ll m_D$ and the use of the low-temperature expansion of the relic potential is justified. At sufficiently high temperatures $T_{\rm cross} \gg m_D$ one must switch to the high-temperature expansion of the relic potential which could lead to zero crossings at earlier times. This possibility is discussed in \cref{sec:conclusions}.

In summary, if 
\begin{equation}
    m_N \lesssim {\rm max}\bigg(\text{\cref{delayed-zero-crossings}, \cref{overclosure}}\bigg)~. 
\end{equation}
then the scalar field's energy density does not overclose the universe.

\subsubsection*{Modified Seesaw Relationship \label{sec:modified-seesaw}}

 For cosmologies in which $T_{\rm cross} < T_{\rm now}$, the typical see-saw relationship does not hold in the present epoch i.e., $m_{\nu,\rm now}\neq m_D^2 / m_N$. This is because if $T_{\rm cross} < T_{\rm now}$  then the present day scalar's potential resembles \cref{fig:pica}. The effective Majorana mass is therefore given by $m_N^{\rm eff}(t)=|g\phi(t)-m_N|=g\varphi(t)$ ({\it cf.}~\cref{mass-matrix}) and this can differ substantially from $m_N$ even after time averaging.

\begin{figure}
    \centering
    \includegraphics[keepaspectratio, width=0.75\textwidth]{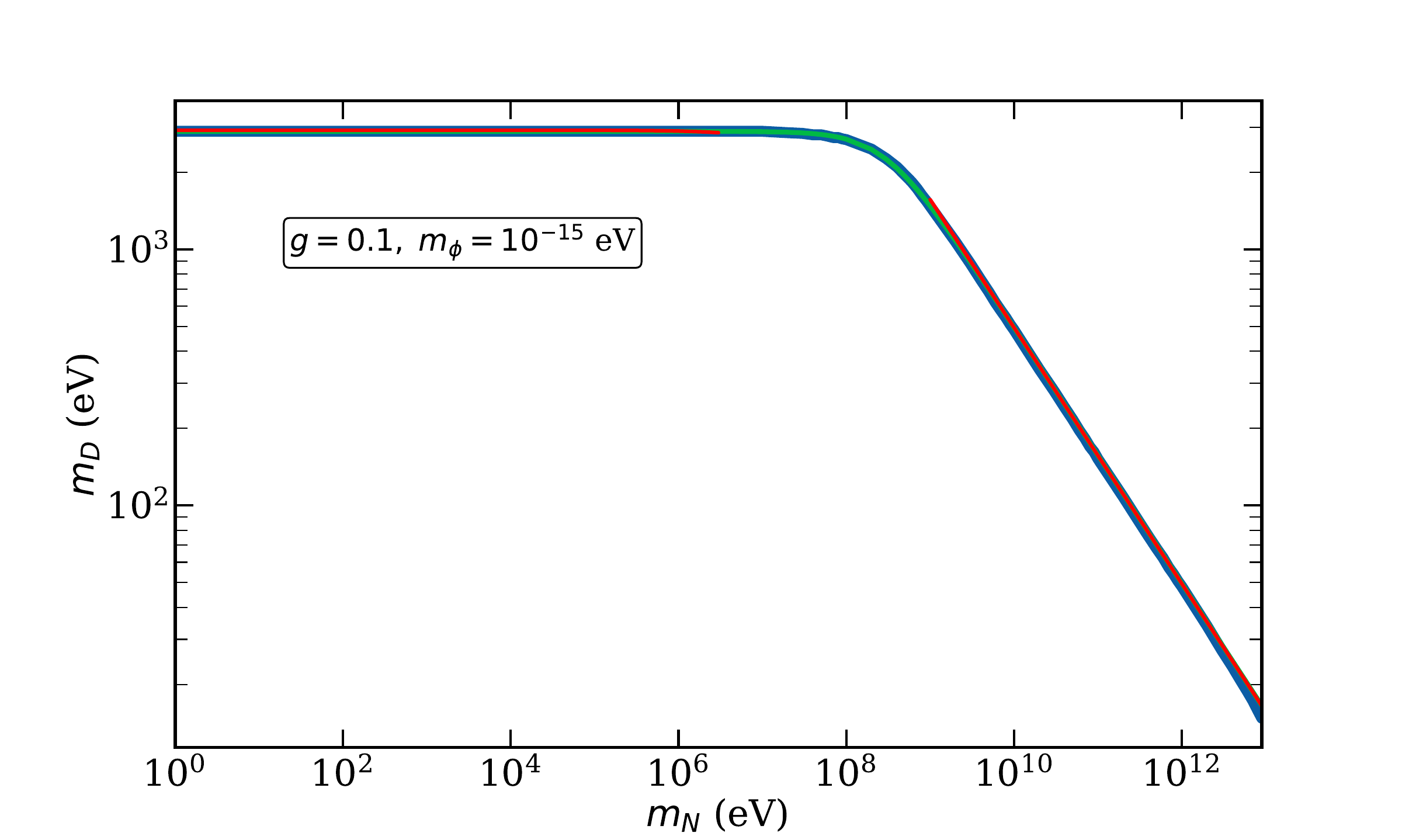}
    \caption{Modified seesaw relationship between $m_D$ and $m_N$ for initial conditions such that $\phi_0>\phi_c$. The curve is defined by demanding $\langle m \rangle_\nu = 0.1 ~ \rm{eV}$ today, and solving for $m_D$. The blue line is the numerical solution, while the green line is the approximation given in \cref{functional-approximation}. For this value of $g$ and $m_\phi$, we find that the parameters are given by $a = 2.9 \times 10^3 ~ \rm{eV}$, $b = 3.4 \times 10^{-9} ~ \rm{eV}^{-1}$, and $c = 1.4$. The red lines are given by $1.6 \times \frac{m_D^2}{g \varphi_{\rm min}}$ for small values of $m_N$ and $4.7 \times \frac{m_D^2}{g \varphi_{\rm min}}$ for large values of $m_N$. We find that these are very good approximations to the numerical solution across different values of $g$ and $m_\phi$ except along the `kink' in the plots. The location of the kink scales inversely with $m_\phi$, i.e. for this value of g it is given by approximately $10^8 \times \frac{10^{-15}}{m_\phi} ~ \rm{eV}$. Notice that $m_D$ actually \textit{decreases} with increasing $m_N$.}
    \label{fig:modified-seesaw}
\end{figure}

We find that $m_D^2/|g\varphi_{\rm min}|$ provides a reasonable parametric estimate for the time-averaged value of the neutrino mass $\langle m_{\nu}\rangle_{\rm now}$ for most of the parameter space we consider (see \cref{fig:modified-seesaw} for an illustration).

In \cref{fig:modified-seesaw} we show the relationship between $m_D$ and $m_N$ obtained by numerically solving for 
\begin{equation}
\label{m_nu-today-condition}
    \langle m_\nu \rangle_{\rm  now} = \frac{\int \dd \chi~m_D \lambda_-(\chi)[V_{\rm max} - V(\chi)]^{-1/2}}{\int \dd \chi ~ \qty[V_{\rm max} - V(\chi)]^{-1/2}} = 0.1 ~ \rm{eV}~,
\end{equation}
over a period of oscillation, with $\lambda_-$ given by \cref{lambda}.\!\footnote{In practice, we use the large-$\chi$ expansion of $\lambda_-$ for the purposes of numerical stability.} We find that the function 
\begin{equation}
\label{functional-approximation}
    m_D = \frac{a}{\left(1 + (b \times m_N\right)^{c})^{\tfrac1{2c}}}~,
\end{equation}
provides a good model for $m_D(m_N)$ when its parameters are fit to the numerical solutions. The parameters $a$, $b$, and $c$ are then functions of $g$ and $m_\phi$. Notice, remarkably, that as $m_N$ increases we always find that $m_D$ must {\it decrease} in order to maintain $\langle m_{\nu,\rm now} \rangle=0.1~{\rm eV}$.

\subsection{Oscillations begin before neutrino decoupling \label{sec:higher-temps}}

So far we have focused on $m_\phi\lesssim 10^{-14} ~{\rm eV}$ such that neutrinos are decoupled relics throughout the entirety of the scalar fields' oscillating dynamics.  For $m_\phi \gtrsim 10^{-14}~{\rm eV}$ one has $T_{\rm osc} \gtrsim T_{\nu,\rm dec}$ and neutrinos are in thermal equilibrium at early times. Eventually, neutrinos will still decouple, and the cosmological epochs described above will follow similar dynamics. Therefore, the constraints derived by considering the cosmological epoch in which $T<T_{\rm dec}$ apply to all cosmologies. Nevertheless, it is interesting to understand how the scalar field's dynamics differs in a cosmological epoch where neutrinos are in thermal and chemical equilibrium.

We first begin by discussing the thermal potential, which assumes that neutrinos scatter with the rest of the Standard Model plasma quickly enough to remain in equilibrium. Then we discuss dynamical implications. At temperatures below a GeV (or so) we find that the neutrino interaction rate is not fast enough to keep neutrinos in thermal equilibrium on (very short) dynamical time scales that determine whether or not the field experiences zero crossings. This modifies the interpretation of certain cosmological constraints as we will see in \cref{sec:m_nu_CMB}. As we will argue in what follows, this inhibits zero-crossings of the field, and we do not expect any strong constraints from the epoch of BBN.

\subsubsection*{The thermal potential}
If oscillations of the $\phi$ field start before neutrino decoupling, then the neutrino is still in the thermal bath. This leads to an additional contribution to the potential from the free energy of the light neutrino of the form \cite{Quiros:1999jp}
\begin{equation}
    \label{V_thermal}
    V_\text{thermal}(\phi) = -\frac{T^4}{2 \pi^2} J_F[m_L^2(\phi) / T^2]~,
\end{equation}
where $J_F$ is the fermionic thermal function (free energy density)
\begin{equation}
    J_F\qty(m_L^2 / T^2) = \int_0^\infty \dd x ~x^2 \log \left(1 + e^{-\sqrt{x^2 + m_L^2 / T^2}}\right)~.
\end{equation}
Here $m_L=m_L(\phi)$ is the mass of the light neutrino species. 
For $T \gg m_L$ the thermal function has a simple high-temperature expansion, 
\begin{equation}
    J_F\qty(m_L^2 / T^2) \simeq \frac{\pi^2}{24} \frac{m_L^2}{T^2} + O(m_L^4/T^4)~. 
\end{equation}
We will assume that the number density of the heavy neutrino is Boltzmann suppressed i.e., that $m_H \gg T$.
\begin{figure}[t]
    \centering
    \begin{subfigure}{0.475\textwidth} 
        \centering
        \includegraphics[width=\textwidth]{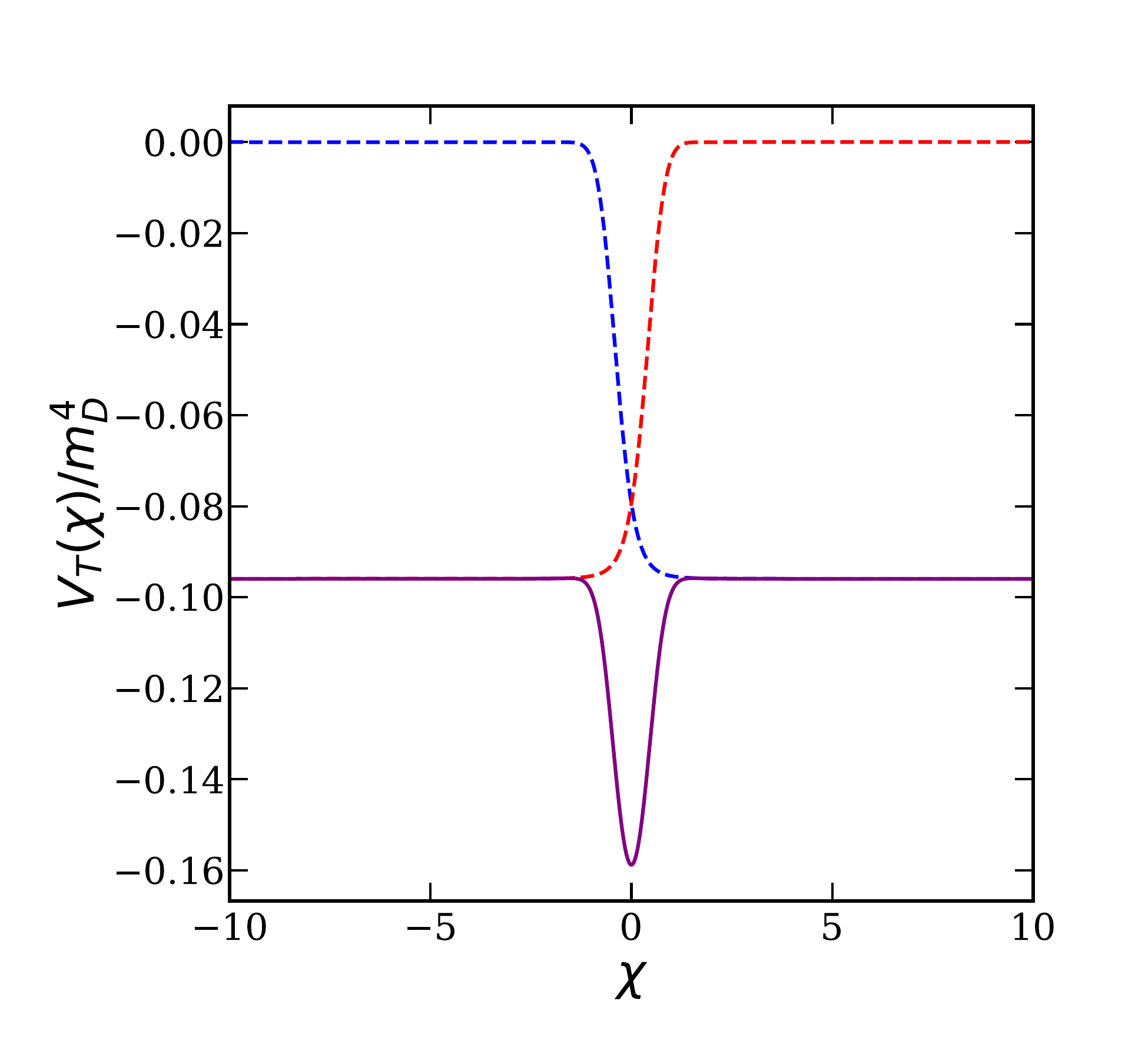}
        \caption{$T=m_D$}
        \label{fig:polaroida}
    \end{subfigure}
    \begin{subfigure}{0.475\textwidth}
        \centering
        \includegraphics[width=\textwidth]{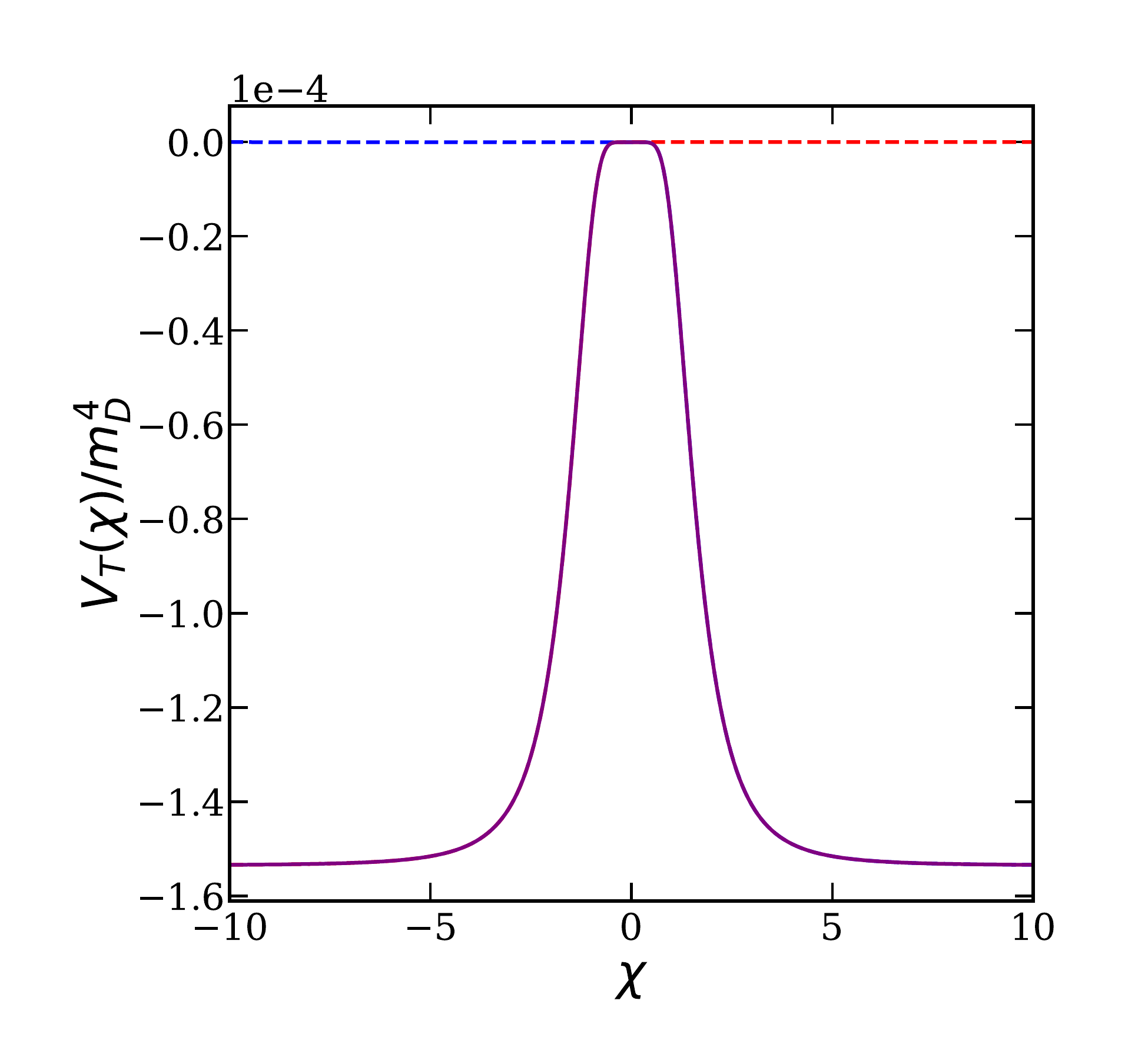}
         \caption{$T=m_D/5 $}
         \label{fig:polaroidb}   
    \end{subfigure}
\caption{The thermal potentials $V_T(\chi)/m_D^4$, as given in \cref{V_thermal}, for $\nu_+$ (red-dashed) and $\nu_-$ (blue-dashed) and their total sum (purple-solid) plotted for different values of $T/m_D$. \textbf{(a)} The potential has a minimum at $\chi=0$ for $T\gtrsim 0.6 m_D$ whereas \textbf{(b)} it has a barrier for $T\lesssim 0.6 m_D$.  }
\label{fig:thermal}
\end{figure}

\subsubsection*{Dynamical implications}

When neutrinos are in thermal equilibrium, the relic potential introduced above is replaced by the thermal potential of \cref{V_thermal}. The thermal potential for $\chi$ stems from the free energy of the neutrino gas (in thermal and chemical equilibrium) as a function of neutrino mass.  Since $m_L$ and $m_H$ are functions of $\phi$, the thermal functions for both the light- and heavy-species are implicit functions of $\phi$, and serve as effective potentials which control the scalar field's dynamics.  For simplicity, we will specialize to initial conditions such that $\phi_0<\phi_c$, however our qualitative conclusions apply to both initial conditions to the left and right of $\phi_c$. 

As shown in \cref{fig:thermal}, depending on the ratio of $m_D/T$ the combined thermal potential for $\nu_+$ and $\nu_-$ can have a minimum or maximum at $\chi=0$. The reliability of the potential, however, depends on whether or not neutrinos have enough time to thermalize. The typical period of an oscillation cycle is set by $\tau_\phi\sim 1/m_\phi$ involving an oscillation amplitude of order $\chi_L$ (or $\chi_R$ when considering initial conditions where $\phi_0>\phi_c$). The sharp features in \cref{fig:thermal} only occur over a region of $\Delta\chi \sim O(1)$. In order for the neutrinos to maintain thermal equilibrium during this small window, they must scatter efficiently over a time scale given by $\Delta\tau\sim (\Delta \chi/\chi_L)\times \tau_\phi$. This is a much shorter time scale than $\tau_\phi$.

Setting the neutrino interaction time-scale, $\tau_\nu \sim 1/(G_F^2 T^5)$ where $G_F=1.116 ~\times 10^{-5}{\rm GeV}^{-2}$ is the Fermi-constant, equal to the relevant time scale $\Delta\tau \sim \tau_\phi/\chi_L$, we find the temperature at which equilibrium can be maintained dynamically 
\begin{equation}
    T_{\rm dyn-eq} \sim 100~{\rm GeV} \qty(\frac{g}{0.1})^{2/7} \qty(\frac{1~{\rm MeV}}{m_D})^{2/7} ~. 
\end{equation}

For $T\lesssim T_{\rm dyn-eq}$ the thermal potential is a bad approximation near $\chi=0$. Instead it is more appropriate to use the relic potential at these temperatures, because the (suddenly) heavy neutrinos do not have enough time to adjust their phase space distribution via scattering and decays. The analysis of \cref{sec:large-negative} therefore applies immediately for $T\lesssim T_{\rm dyn-eq}$, and we expect no zero crossings.

Since $T_{\rm dyn-eq} \gg T_{\rm BBN}$ we expect any effects due to the thermal potential (e.g., spectral distortions or decays $\nu_H \rightarrow \nu_L \phi$) to occur at early epochs and to be diluted by cosmic expansion. Furthermore, since  $T_{\rm dyn-eq}$ depends weakly on both $g$ and $m_D$, we do not expect any interesting constraints from the epoch of BBN (e.g., the helium to deuterium abundance ratio $Y_p$) for any of the parameter space we consider.

\section{Phenomenology \label{sec:pheno} }
In this section we discuss a number of different constraints that arise both from laboratory experiments and cosmological considerations. The constraints are summarized in \cref{sec:summary_plots} in \cref{fig:parameter_space_left,fig:parameter_space_right} (see also \cref{rup-cond} which is not plotted).

\subsection{Neutrino oscillation experiments \label{sec:dinos}}
The time evolution of cosmic fields, such as our ultralight scalar dark matter candidate, can distort measurements of neutrino oscillation parameters. When $\tau_\phi$ is very long, explicitly time varying effects (such as periodicities in the solar neutrino flux) can be searched for using solar neutrinos \cite{Berlin:2016bdv}. If the period of oscillation, $\tau_{\phi}=2\pi/m_\phi$, is  large relative to neutrino time of flight, $t_{\nu, \rm tof}$, but small compared to the run-time of an experiment, DiNO signals are observable.  For the smallest values of $\tau_\phi$ (or the largest values of $m_\phi$), oscillations are too fast and DiNO signals cannot probe these regions of parameter space ($m_\phi\gtrsim 10^{-11}~{\rm eV}$).

We follow \cite{Krnjaic:2017zlz} and set constraints at constant $\eta_\phi$, as defined using the explicit model given by \cref{model-def}, as
\begin{equation}
\label{kamland}
    \eta_\phi \equiv \frac{\delta m_L }{m_L} ~, 
\end{equation}
For \cref{fig:parameter_space_left} we set a constraint at $\eta_\phi\lesssim 0.032$  from the KamLAND constraint of \cite{Krnjaic:2017zlz}, which is the most restrictive DiNO constraint for $m_\phi\lesssim 10^{-11}~{\rm eV}$, and $\eta_\phi\lesssim 0.21$ for $10^{-11}~{\rm eV} \lesssim m_\phi \lesssim 10^{-9}~{\rm eV}$ from the Daya Bay constraint. For $T_{\rm drop} > T_{\rm now}$, where the brief period of anharmonicity in oscillations to the left of $\phi_c$ has ended, we have relaxed to approximately harmonic oscillations about the minimum of the bare potential (see \cref{fig:imgc}). Therefore in this regime, $\eta_\phi = g\sqrt{2 \rho_{\rm DM,now}^\oplus}/(m_\phi m_N)$ is used to calculate the constraint on $m_N$ (just like in \cite{Krnjaic:2017zlz}) where $\rho_{\rm DM,now}^\oplus$ is the local density of dark matter, $2 \times 10^{-6} \rm ~ eV$ . Below this line, the `drop' has not yet occurred today and the oscillations of the scalar field are very large (see \cref{fig:imga}). As such, we expect neutrino parameters, such as the mass, to vary substantially (i.e., with an $O(1)$ oscillation amplitude) in time and be constrained by DiNO measurements for $m_\phi \lesssim 10^{-9}~{\rm eV}$ for Daya Bay, and $m_\phi \lesssim 10^{-11}~{\rm eV}$ for KamLAND \cite{Krnjaic:2017zlz,Brdar:2017kbt}.

For \cref{fig:parameter_space_right} the correct interpretation of the constraints on $\eta_\phi$ is more subtle. Unlike in Refs.~\cite{Krnjaic:2017zlz,Brdar:2017kbt}, we do not assume sinusoidal oscillations about the potential's minimum, nor do we assume the standard see-saw relationship between $m_N$ and $m_D$. Instead $m_D$ is related to $m_N$ via the modified seesaw relationship approximated by \cref{functional-approximation}. Based on \cref{kamland}, as a rough proxy for these experimental constraints we calculate 
\begin{equation}
    \eta_\phi \sim \frac{\mu_-(\varphi_L) - \mu_-(\varphi_R)}{m_{\nu, \rm now}} = \frac{\mu_-(\varphi_L) - \mu_-(\varphi_R)}{0.1 ~ \rm{eV}}~,
\end{equation}
with $\mu_-$ given by \cref{eigenvalues}. The left and right turning points, $\varphi_{L/R}$ (or equivalently, $\chi_{L/R}$), are determined by numerically solving \cref{turning-point-condition} and using \cref{functional-approximation} to determine $m_D$ for a given $m_N$, $m_\phi$, and $g$. We find that this approximation for $\eta_\phi$ is consistently around 80 - 90 for $T_{\rm cross} < T_{\rm now}$, which therefore rules this region of parameter space out (the rest of parameter space being mostly ruled out by overclosure, see \cref{fig:parameter_space_right}). 

To perform the full analysis for \cref{fig:parameter_space_left} would require a similar determination of the modified seesaw relationship in the parameter space where $T_{\rm drop}$ has not yet occurred and oscillations are asymmetric. One could then follow a similar procedure as described above to calculate \cref{kamland} and estimate if this region of parameter space is  ruled-out by DiNO experiments. A complete analysis would compute $m_L(t)$ properly solving the equations of motion, but this lies beyond the scope of this paper.  

\subsection{Neutrino mass and the CMB \label{sec:m_nu_CMB} }
As mentioned in the introduction, previous studies of ULDM coupled to neutrinos have inferred a bound from the cosmic neutrino background. This is obtained by demanding $\langle m_\nu\rangle _{\rm RMS} \lesssim 0.1~{\rm eV}$ at CMB where the right-hand side coming from published constraints from the Planck collaboration.\!\footnote{The constraint is on the sum of neutrino masses but since we work in a simplified $1+1$ framework this is irrelevant for our present discussion.} In \cite{Krnjaic:2017zlz} this translates into a constraint on $\eta_\phi >(9\times 10^{-3}~{\rm or}~0.1)$ being excluded (depending on whether couplings to lighter or heavier active neutrinos are considered). As we will now argue, this constraint can be substantially modified once the relic potential is incorporated. 

\subsubsection*{Scenarios where naive constraints applies}
Let us begin by discussing which scenarios map onto the analysis in \cite{Krnjaic:2017zlz} most closely. When \cref{boring-inequality} is satisfied, the relic potential is not important at the CMB. This allows the naive constraint, where the field is redshifted with the typical $\rho \propto (T_{\rm now}/T)^3$ scaling, to apply immediately. Notice, however, that the parameter space where \cref{boring-inequality} applies gives $\eta_\phi\ll 10^{-3}$ and so the CMB constraint on the neutrino mass is not informative. We therefore conclude that the CMB never supplies an interesting constraint on the model in the regions where the neutrino background density can be ignored for all $T<T_{\rm osc}$. 

\subsubsection*{Large negative initial conditions \label{sec:neg-pheno}}

Next, consider oscillations that begin to the left of $\phi_c$. If  $T_{\rm drop}\gg T_{\rm CMB}$ then the oscillations are predominantly about $\phi_{\rm min}$, and therefore well separated from $\phi_c$; in this case the analysis of \cite{Krnjaic:2017zlz} again apply immediately. 

If $T_{\rm drop} \lesssim T_{\rm CMB}$, then the amplitude of oscillations at the epoch of the CMB is large. The light neutrinos, $\nu_L$, spend part of their time with $|\chi|\sim O(1)$ which leads ``large'' masses, $m_L \sim m_D$, on the order of the Dirac mass. The time in which $m_L\sim m_D$ is short, being suppressed by $\Delta \tau /\tau_\phi \sim 1/|\chi_L|$, where $\Delta \tau$ is the time spent near the turning point and $\tau_\phi$ is the period of oscillation. Therefore we will consider parameter space such that  
\begin{align}
    T_{\rm now}\lesssim T_{\rm drop} \lesssim T_{\rm CMB}~, 
\end{align}
corresponding to the green band in \cref{fig:parameter_space_left}. By focusing on regions where $T_{\rm now} < T_{\rm drop}$ we are able to use the typical see-saw relationship to relate $m_D$ and $m_N$.  With these assumptions, in order for the mass of neutrinos to be sufficiently small at the epoch of the CMB, we require $m_D/\chi_L \lesssim 0.1~{\rm eV}$, or equivalently (using $\chi_L \simeq g \phi_L/(2 m_D)$), 
\begin{equation}
\label{m_nu_bound_left}
    \frac{2m_D^2 m_\phi}{g\sqrt{2 \rho_{\rm DM, now}} }  \qty(\frac{1}{1100})^{3/2}  \lesssim 0.1~{\rm eV} ~, 
\end{equation}
to avoid constraints from the CMB, where we use $m_\nu \lesssim 0.1~{\rm eV}$ as a rough proxy for the constraint on the sum of neutrino masses.

We find that \cref{m_nu_bound_left} is a reasonable approximation to the numerical solution (obtained by solving \cref{phi-eom} for a period of oscillation, neglecting Hubble friction), but typically overestimates the time-averaged neutrino mass by a factor of $\sim 2$.  When calculating the numerical solution we solve, $\ddot{\chi} = -\frac{dV}{d \chi}$, with $V(\chi)$ given by \cref{chi-potential-left} where we take $T \ll m_D \lambda_+$ at $T_{\rm CMB}$. \!\footnote{In practice, when calculating the mass we use the large-$|\chi|$ expansion of $m_D \lambda_+$ for the purposes of numerical stability.} We then use the Taylor-expansion of $m_D \lambda_+$ for large $|\chi|$ with $\lambda_+$ given by \cref{lambda} to determine the time-average of the mass over the oscillation period, which we demand is less than $0.1~{\rm eV}$. This gives us an upper bound on $m_D$ in terms of $m_\phi$. 

Translating \cref{m_nu_bound_left} from $m_D$ to $m_N$ using the seesaw relationship (that applies today) $m_{\nu, \rm now} = 0.1 ~ {\rm eV} = m_D^2 / m_N$, we obtain a bound on $m_N$ above which neutrinos are too heavy at the epoch of the CMB (blue region in \cref{fig:parameter_space_left}). However, for sufficiently high $m_N$, the constraint no longer applies. This can be understood in the context of \cref{fig:imgc}. After the ``drop'', the potential takes the approximate shape of a simple harmonic oscillator with minimum given by (negative) \cref{psi_c}. The minimum moves further away from $\varphi = 0$ in \cref{fig:eigenvalue_plot} for increasing $m_N$ which tends to decrease the value of $\langle m_\nu \rangle_{\rm CMB}$. Eventually, the minimum becomes so well-separated from the relic potential wall that we can neglect its contribution entirely and treat the potential as just the bare, harmonic potential. In this regime, the time-average of the mass-oscillations are well-approximated by their value at the minimum, which necessarily matches $m_\nu$ today. 

\begin{figure}[!t]
\centering
    \includegraphics[keepaspectratio, width=0.75\textwidth]{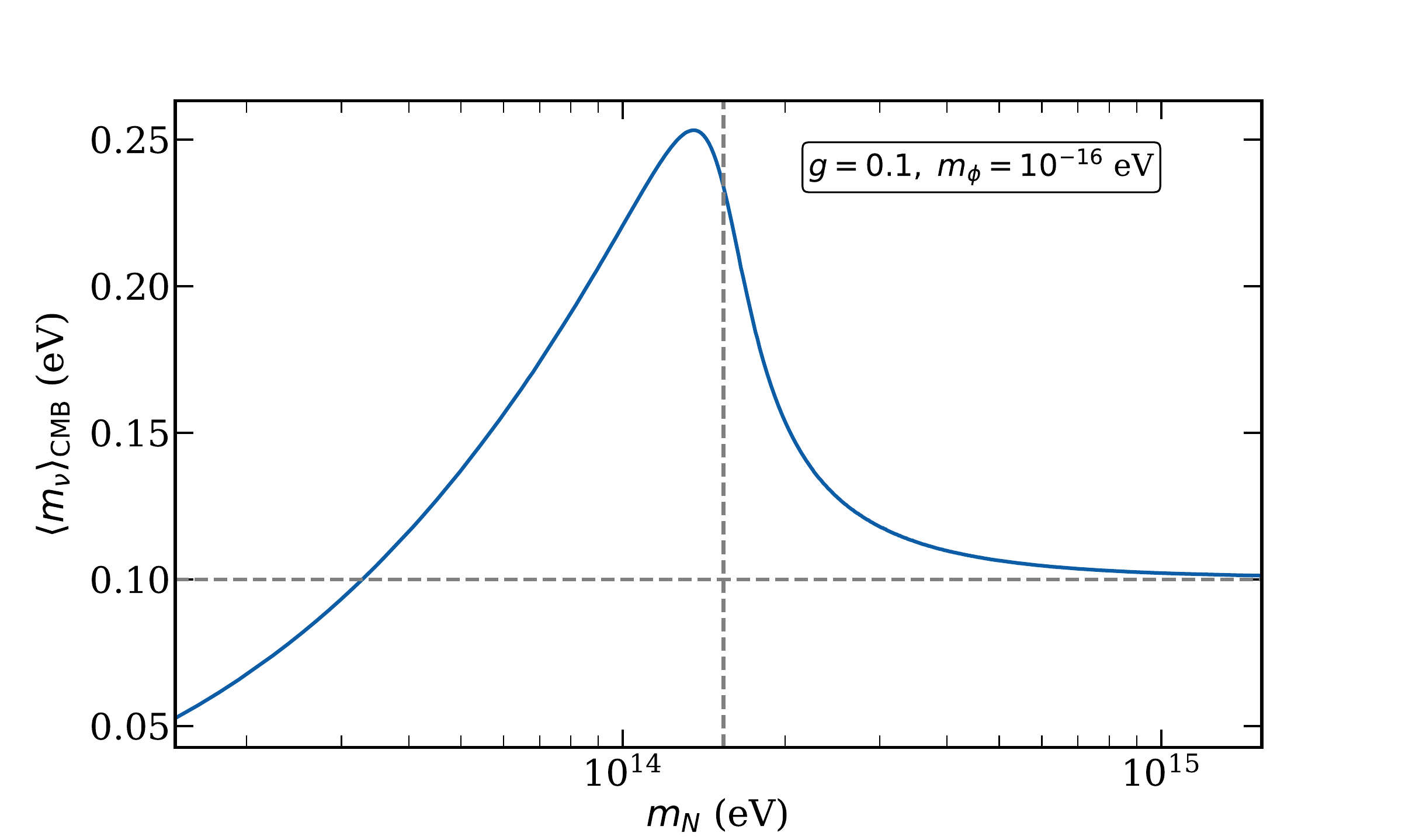}
\caption{Time-averaged neutrino masss at the epoch of the CMB, $\langle m_\nu \rangle_{\rm CMB}$ as a function of $m_N$, for initial conditions to the left. We have used $g = 0.1$ and $m_\phi = 10^{-16} ~ \rm{eV}$ as illustrative parameters. The vertical dashed gray line represents the value of $m_N$ for which $T_{\rm drop} = T_{\rm CMB}$, while the horizontal dashed gray line represents $\langle m_\nu \rangle_{\rm now}$. Notice that for sufficiently high $m_N$, $\langle m_\nu \rangle_{\rm CMB}$ asymptotically approaches $\langle m_\nu \rangle_{\rm now}$.
\label{fig:avg_mass_plot} }
\end{figure}

This behavior is shown in \cref{fig:avg_mass_plot}, where one can clearly see the crossover between the two different regimes. The cross over occurs at $m_N$ such that $T_{\rm drop}(m_N) \simeq T_{\rm CMB}$ (obtained by solving \cref{T_drop}). As $m_N$ tends to larger values the average neutrino mass at $T_{\rm CMB}$ becomes equal to its value now, which we have fixed to be $0.1 ~{\rm eV}$.

\subsubsection*{Large positive initial conditions}
Next we consider the case of oscillations to the right of $\phi_c$ with $T_{\rm cross} < T_{\rm now}$ such that zero crossings have never occurred. In this regime the minimum of the potential tends closer to the steep relic potential wall at $\chi = 0$ in \cref{fig:pica} as the temperature decreases (see \cref{phi0_right:chi_min_big,phi0_right:chi_min_small}). This means that the average neutrino mass tends to increase with decreasing temperature. Therefore we are guaranteed that if neutrinos are $0.1~{\rm eV}$ today, then they were lighter than $0.1~{\rm eV}$ at the CMB and no constraint exists. For $T_{\rm cross} > T_{\rm now}$ the parameter space is mostly ruled out by overclosure and so we do not consider this regime here (see \cref{fig:parameter_space_right}).

\subsection{Supernova cooling \label{sec:SN-cooling}}
Constraints from supernova cooling were analyzed in Ref.~\cite{Krnjaic:2017zlz}. The assumed energy loss channel was $\nu_L \bar{\nu}_L \rightarrow \phi \phi$ with a cross section given parametrically by 
\begin{equation}
    \sigma \sim \frac{g^4}{T^2} \qty(\frac{m_L}{m_H})^4~. 
\end{equation}
This scaling stems from the need to insert two mixing angles $\theta_L^2\sim (m_L/m_H)$ in the leading Feynman diagram; the temperature scaling arises from naive dimensional analysis. 

In the interior of a supernova the density of neutrinos is very large (just like in the early universe). An immediate consequence is that mostly-active neutrinos are much lighter in a supernova than in vacuum $m_{L,{\rm SN} } \ll m_{L, {\rm vac}}$, while mostly-sterile ones are much heavier $m_{H, {\rm SN}} \gg m_{H, {\rm vac}}$. For $T_{\rm SN}\sim 30~{\rm MeV}$ this suppresses the cross section for $\nu \bar{\nu} \rightarrow \phi \phi$ by orders of magnitude and essentially eliminates the supernova cooling constraint considered in Ref.~\cite{Krnjaic:2017zlz}. 

Light neutrinos can also annihilate via an off-shell heavy neutrino with a cross section that scales, for $T\ll m_H$, as 
\begin{equation}
    \label{sigma-contact}
    \sigma \sim \frac{g^4}{m_H^2} \qty(\frac{m_L}{m_H})^2 \gg \frac{g^4}{T^2} \qty(\frac{m_L}{m_H})^4~. 
\end{equation}
We therefore conclude that for any temperature satisfying $T\gg m_{L}$ 
~\\
~\\
\begin{equation*}
      \begin{fmffile}{big}
    \begin{fmfgraph*}(75,75) 
    \fmfstraight
    \fmfleft{i1,i2}
    \fmfright{o1,o2}
    \fmf{fermion}{i1,v1}
    \fmf{fermion,label=$\nu_H$}{v1,v2}
    \fmf{fermion}{v2,i2}
    \fmf{dashes}{v1,o1}
    \fmf{dashes}{v2,o2}
    \fmfv{label=$\nu_L$}{i1}
    \fmfv{label=$\nu_L$}{i2}
    \fmfv{label=$\phi$}{o1}
    \fmfv{label=$\phi$}{o2}
\end{fmfgraph*}
\end{fmffile} 
\hspace{0.1\linewidth}
\raisebox{37.5pt}{$\gg$}
\hspace{0.1\linewidth}
  \begin{fmffile}{small}
    \begin{fmfgraph*}(75,75) 
    \fmfstraight
    \fmfleft{i1,i2}
    \fmfright{o1,o2}
    \fmf{fermion}{i1,v1}
    \fmf{fermion,label=$\nu_L$}{v1,v2}
    \fmf{fermion}{v2,i2}
    \fmf{dashes}{v1,o1}
    \fmf{dashes}{v2,o2}
    \fmfv{label=$\nu_L$}{i1}
    \fmfv{label=$\nu_L$}{i2}
    \fmfv{label=$\phi$}{o1}
    \fmfv{label=$\phi$}{o2}
\end{fmfgraph*}
\end{fmffile} 
\raisebox{37.5pt}{\quad\quad\quad.}
\end{equation*}
~\\
We will follow the same parametric estimate sketched in Ref.~\cite{Krnjaic:2017zlz}, but using the $\nu^2\phi^2$ (that arises from integrating out $\nu_H$) vertex rather than the $\nu^2\phi$ vertex. Taking the rate of energy loss per neutrino per unit time to be $\dot{E}_\nu \sim T_{\rm SN} n\sigma$, and a thermal density of $n_\nu \sim 9\zeta(3)T_{\rm SN}^3/(4\pi)^2$, 
we find
\begin{equation}
    \dot{E}_\nu\sim 5 \times 10^{-4}~\frac{{\rm erg}}{s} \qty(\frac{T_{\rm SN}}{30~{\rm MeV}})^4 \qty(\frac{1~{\rm GeV}}{m_H})^4 \qty(\frac{m_L(T_{\rm SN})}{0.1~{\rm eV}})^2\qty(\frac{g}{0.1})^4~. 
\end{equation}
For $\sim 4\times 10^{54}$ neutrinos in a supernova, we require $\dot{E}_\nu \lesssim 10^{-5} {\rm erg}/{\rm s}$. Since $m_L(T\sim 30~{\rm MeV}) \ll m_{\nu,~{\rm now}}$  and $m_H$ is generically much larger than $m_N$, we expect that supernova bounds are essentially always satisfied. We therefore do not include them in our summary plots \cref{fig:parameter_space_left,fig:parameter_space_right}.

\subsection{Cosmologically stable condensate}
In order for the misaligned scalar condensate to survive at late times we require that particles in the bath do not destroy the condensate. In particular the scattering of neutrinos from the condensate $\nu \phi \rightarrow \nu \phi$ should have a rate smaller than Hubble $H$ \cite{Dymnikova:2000gnk}. 

As discussed in \cref{sigma-contact}, the largest scattering rate (per $\phi$ particle) is mediated by an off-shell heavy neutrino. A conservative requirement is that this rate is small compared to Hubble at all cosmic epochs after reheating. In what follows we will assume that $T_{\rm RH} \gtrsim 10~{\rm MeV}$ (slightly larger than the estimated minimum temperature of $4~{\rm MeV}$). 

The rate per scalar particle (in the $m_L\ll m_H$ limit) scales as 
\begin{equation}
    \Gamma \sim \frac{g^2 T^3}{(4\pi) m_H^2} \theta_L^4 ~, 
\end{equation}
The scaling of $m_L(T)$ and $m_H(T)$ with temperature depend on the initial conditions. For initial condition that begin at large negative field values, the mixing angle is maximized at the right turning point $\chi=\chi_R$ (the closest point of approach to a Dirac pair). 

\subsubsection*{Oscillations begin after $T_{\rm \nu,dec}$}
For simplicity we will focus on the case of large negative initial conditions for the field $\phi$. The mixing angle suppression of of the $\nu \phi \rightarrow \nu \phi$ cross section, $\sin^4\theta$, is time-dependent. The cross section is largest close to the right-turning point $\phi_R$ given in \cref{no-zero-crossings}. When written in terms of $\chi$ we have 
\begin{equation}
    \chi_R= - \qty(\frac{m_D n_\nu(T) }{2 \rho_{\phi}(T)})~.
\end{equation}
For $m_D \gtrsim 100 m_{\nu,~\rm now} \simeq 10~{\rm eV}$,  $|\chi_R| \gg 1$ and we have 
\begin{equation}
    \sin^2\theta  \simeq \frac{1}{4\chi^2}~, 
\end{equation}
such that (also using the large $\chi$ approximation for $m_H$)
\begin{equation}
    \Gamma \sim \frac{g^2 T^3}{(4\pi)m_D^2}\frac{1}{\chi^8}~. 
\end{equation}
Since $\chi_R$ is independent of temperature until $T\sim T_{\rm drop}$, and the rate is proportional to $T^3$, this rate is maximized at the highest available temperature.

When integrated over an oscillation cycle, the time average is dominated by $\chi \simeq \chi_R$. We find that the rate, averaged over one oscillation cycle, is given by 
\begin{equation}
    \Gamma \sim \frac{g^2 T^3}{(4\pi)m_D^2}\frac{1}{\chi_R^8 \chi_L}~.
\end{equation}
Notice that $\chi_L \propto T^{3/2}$ and therefore $\chi_L\gg1$. Furthermore, even for modest value of $\chi_R$ (e.g., $\chi_R\sim 100$) the suppression from $1/\chi_R^8$ is extremely large. 

If the field has not yet started oscillating, then $\chi$ is stuck at $\chi_L$ and we find a rate that is suppressed by $1/\chi_L^8$. We therefore take the largest possible temperature as $T_{\rm osc}\sim \sqrt{M_{\rm Pl} m_\phi}$.  Demanding $\Gamma \lesssim H \sim T^2/M_{\rm Pl}$ requires, 
\begin{equation}
    \frac{g^2}{\chi_R^8 \chi_L} \ll  \frac{m_D^2}{T_{\rm osc} M_{\rm Pl}}~.
\end{equation}
This condition is easily satisfied even for $g\sim O(1)$. We therefore do not include constraints from a ruptured condensate for $m_\phi\leq 10^{-14}~{\rm eV}$ in what follows.

\subsubsection*{Oscillations begin before $T_{\rm \nu,dec}$}
For larger masses $m_\phi \geq 10^{-14}~{\rm ev}$, the field will oscillate prior to neutrino decoupling. If the field is able to cross $\chi=0$, this can lead to short periods in which there are pseudo-Dirac pairs each coupling to $\phi$ with a strength of $g/\sqrt{2}$. The period of time spent near $\chi\sim O(1)$ is suppressed by the large oscillation amplitude $\chi_L$, such that the time averaged mixing angle is 
\begin{equation}
    \langle \sin^2\theta \rangle \sim \frac{1}{2 \chi_L}~. 
\end{equation}
Therefore, in this epoch, the rate averaged over an oscillation cycle is set by 
\begin{equation}
    \Gamma \sim 2\times \frac{g^2T^3}{(4\pi) m_D^2}  \frac{1}{2 \chi_L} = \frac{g m_\phi T^3  }{(4\pi)m_D \sqrt{2\rho_{\rm DM}}}  ~,
\end{equation}
where the pre-factor of $2$ accounts for both $\nu_{+}$ and $\nu_{-}$ during this epoch. 

Using $\rho_{\phi}\approx 0.7~{\rm eV} \times T^3$ we find 
\begin{equation}
    \Gamma \sim  \frac{g~T^{3/2} m_\phi }{8\pi m_D \sqrt{1.4~{\rm eV}}} ~. 
\end{equation}
Notice that when measured in units of Hubble in a radiation dominated universe, $H\sim T^2/M_{\rm Pl}$, the rate will be fastest at the {\it highest} temperatures. In order for the condensate not to rupture we therefore require that $\Gamma(T_{\rm osc}) < H(T_{\rm osc}) = m_\phi/3$, which leads to 
\begin{equation}
    \label{rupture-1}
     \qty(\frac{g^2T_{\rm osc}^2}{64\pi^2 m_D^2})\qty(\frac{T_{\rm osc}}{1.4~{\rm eV}}) <\frac{1}{9}~, ~\qq{or}~~m_D \gtrsim 10~{\rm MeV}~\qty(\frac{T_{\rm osc}}{1~{\rm MeV}})^{3/2} \qty(\frac{g}{0.1}) ~. 
\end{equation}

Alternatively, we can avoid rupturing the condensate if $m_D\gtrsim 0.6\times T_{\rm osc}$ such that there is a barrier (rather than minimum) near $\chi=0$ as shown in \cref{fig:polaroida,fig:polaroidb}. Therefore, if $g$ is larger than the bound in \cref{rupture-1}, but 
\begin{equation}
    \label{rupture-2}
    m_D\geq 0.6 \times T_{\rm osc}~, 
\end{equation}
then the condensate does not rupture because the thermal potential does not allow the field to cross $\chi=0$. 

In summary, we expect that the condensate will survive provided
\begin{equation}\label{rup-cond}
    m_D \geq {\rm min}\qty(~\text{\cref{rupture-1}~,~\cref{rupture-2}}~)~.
\end{equation}
For $T_{\rm drop} > T_{\rm now}$ this gives a lower-bound on $m_N$ via $m_D = \sqrt{m_{\nu,\rm now} m_N}$ which only applies to already-excluded parameter space in \cref{fig:parameter_space_left}. For $T_{\rm drop} < T_{\rm now}$ we would need to determine the modified seesaw relationship to correctly incorporate the constraint of \cref{rup-cond}, which lies outside the scope of this paper.


\vfill 
\pagebreak 

\subsection{Summary of constraints \label{sec:summary_plots} }

\begin{figure}[H]
\centering
    \includegraphics[keepaspectratio, width=0.85\textwidth]{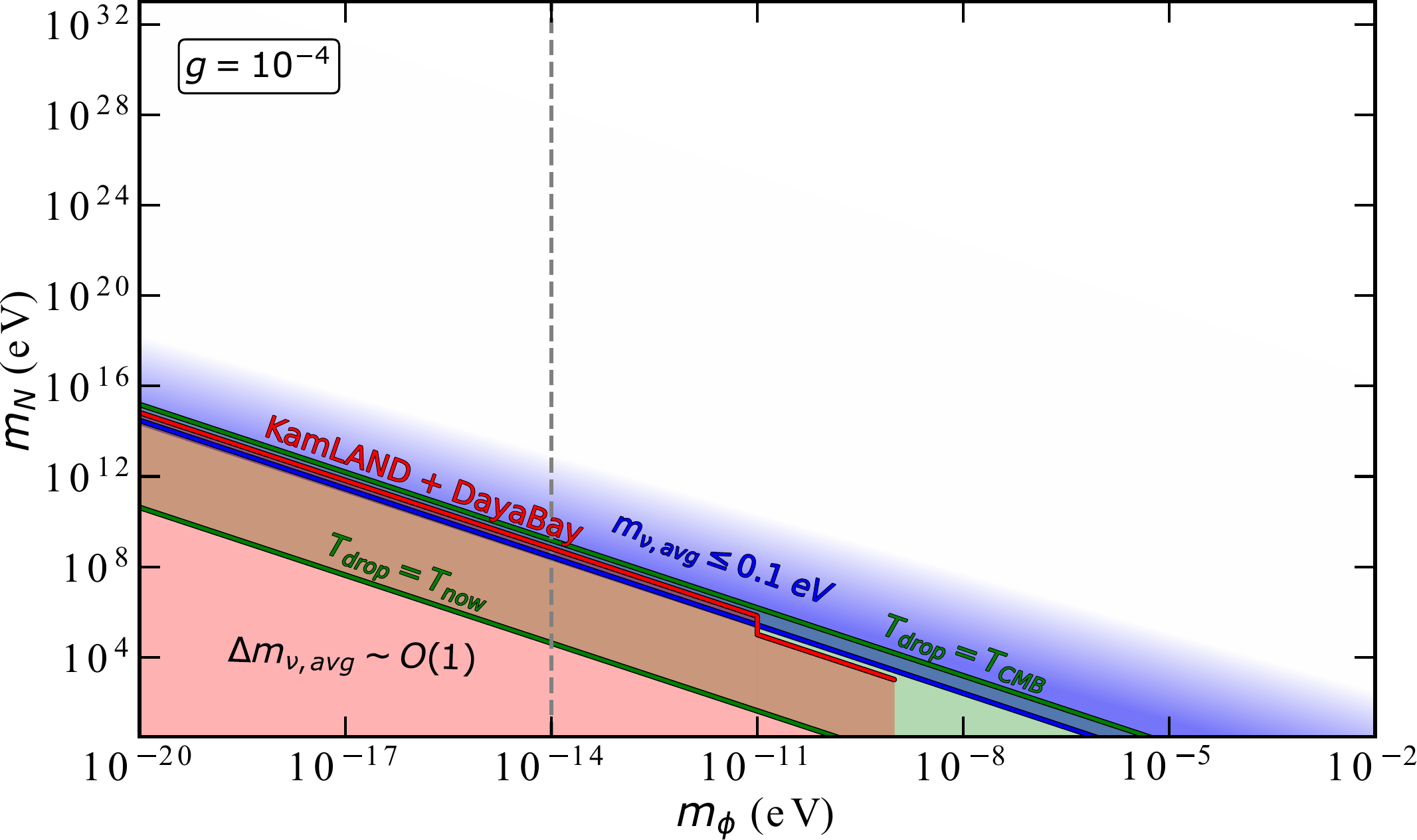}
    \includegraphics[keepaspectratio, width=0.85\textwidth]{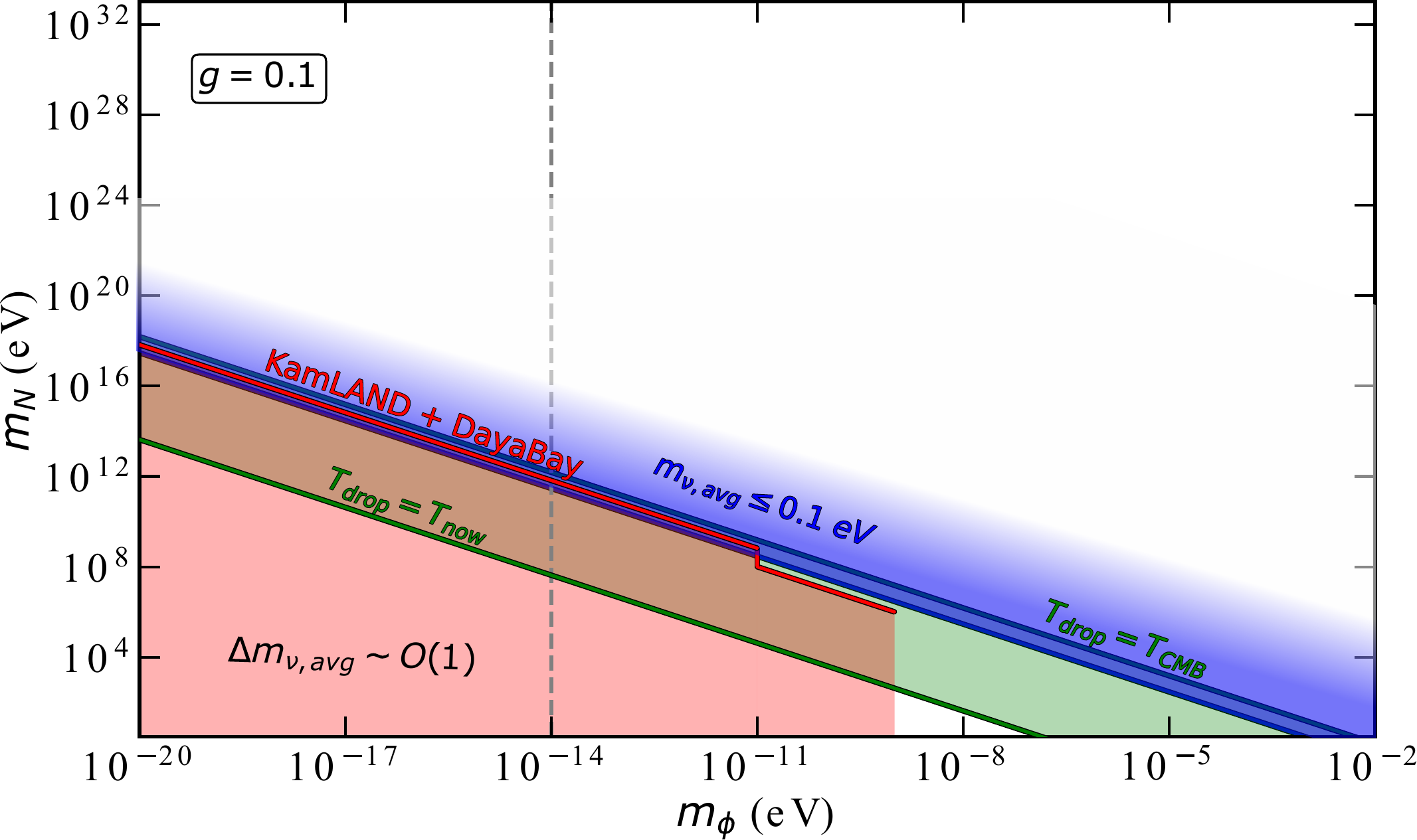}
\caption{Parameter space for $\phi$-induced variations in the neutrino mass for $\phi_0<0$. The gray dashed line marks $m_\phi=10^{-14}~ {\rm eV}$ above which $T_{\rm osc} >T_{\nu, \rm dec}$. The red-shaded region is ruled out by searches for DiNOs (\cref{sec:dinos}). The blue region leads to unacceptably large time-averaged neutrino masses at $T_{\rm CMB}$ (\cref{sec:neg-pheno}); we shade with a gradient since this is a smooth cross-over as shown in \cref{fig:avg_mass_plot}. The green band denotes the region where there is a sudden ``drop'' in the dark matter energy density between the epoch of the CMB and now; this region is disfavoured (see \cref{brick-wall-delta}). Below the green band the scalar  has a large oscillation amplitude today and we expect $O(1)$ time-variation of neutrino mass. \label{fig:parameter_space_left}}
\end{figure}

\begin{figure}[H]
\centering
    \includegraphics[keepaspectratio, width=0.85\textwidth]{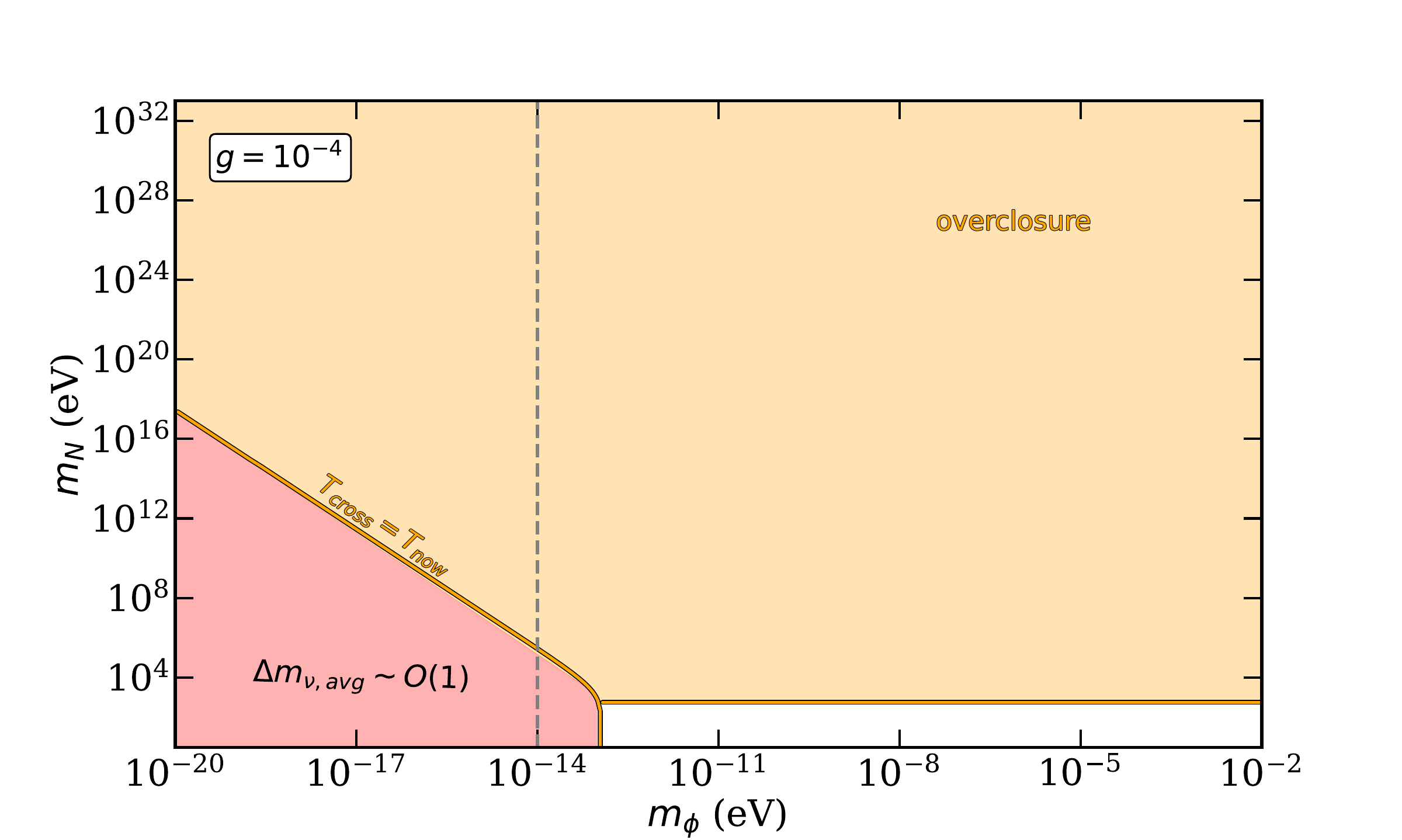}
    \includegraphics[keepaspectratio, width=0.85\textwidth]{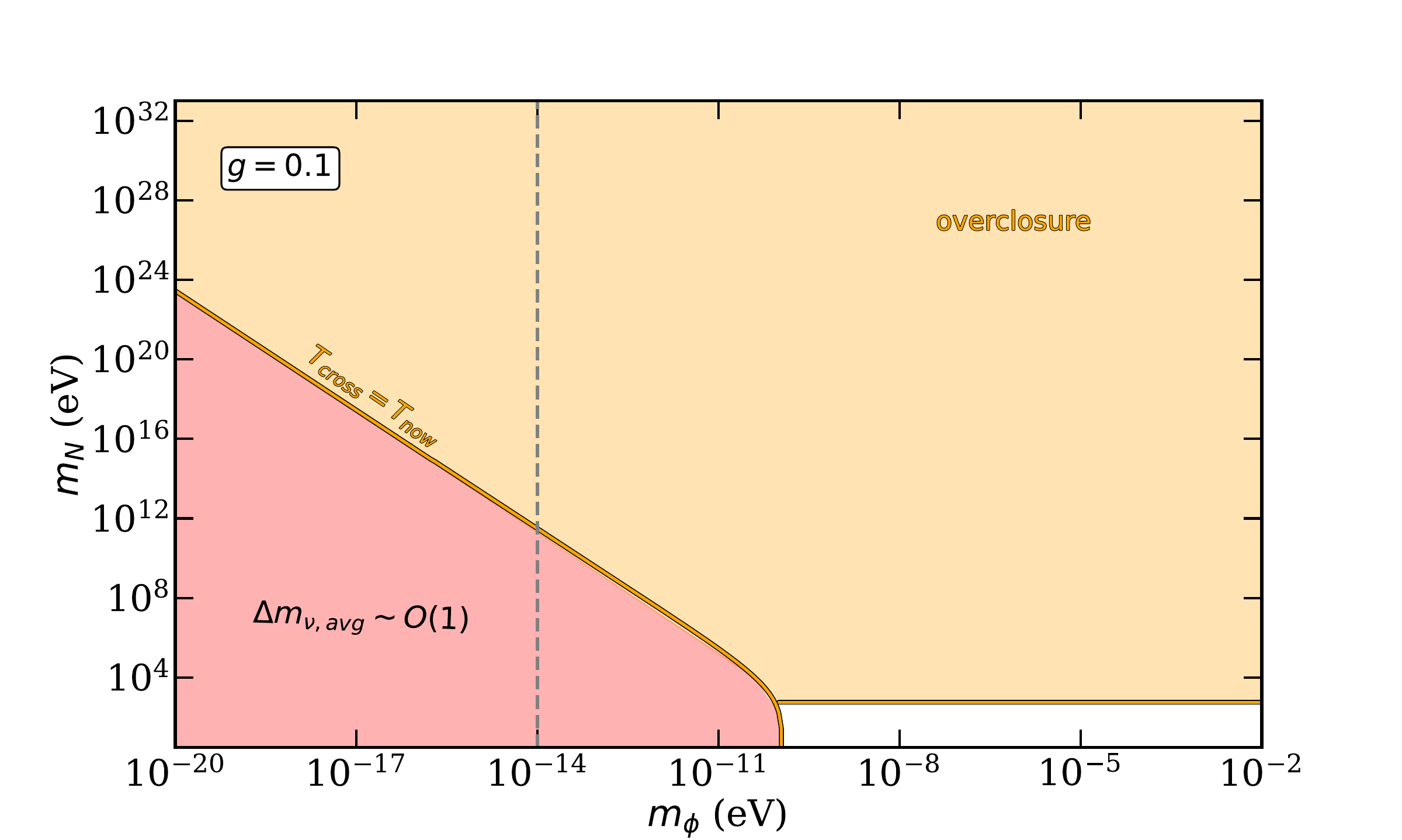}
\caption{Parameter space for $\phi$-induced variations in the neutrino mass for $\phi_0>\phi_c$. The gray dashed line marks $m_\phi=10^{-14} ~ {\rm eV}$ above which $T_{\rm osc} >T_{\nu, \rm dec}$. The red region is ruled out by searches for DiNOs (\cref{sec:dinos}). In the yellow region the field is trapped at $\phi>\phi_c$ at early times, but eventually crosses $\phi_c$ and overcloses the universe (\cref{sec:delayed-zero-crossings}). \label{fig:parameter_space_right}}
\end{figure}

For simplicity we organize our constraints according to initial conditions. 
\Cref{fig:parameter_space_left} shows constraints for initial conditions such that $\phi_0<0$ whereas \cref{fig:parameter_space_right} shows constraints for initial field values satisfying $\phi_0>\phi_c$. 

\pagebreak

As can be seen by comparing \cref{fig:parameter_space_left,fig:parameter_space_right}, we find that the cosmology differs qualitatively depending on whether $\phi_0$ is less than or greater than $\phi_c$. In the former case, \cref{fig:parameter_space_left}, neutrinos retain their adiabatic label throughout all of cosmic history. We find that constraints from $\langle m_\nu \rangle_{\rm CMB}$ complement rather than compete with constraints from DiNOs. When $\phi_0>\phi_c$, \cref{fig:parameter_space_right}, the scalar field either crosses $\phi_c$ before today, or it has not yet crossed $\phi_c$ in which case it is oscillating in a shallow and highly asymmetric potential. The former option is ruled out because it would overclose the universe (unless $m_N\leq 600~{\rm eV}$), while the latter scenario is rule out by DiNOs. 

\pagebreak 

\section{Conclusions and outlook \label{sec:conclusions} }

The main results of this paper pertain to the cosmology of scalar fields coupled to multiple fermions. The eigenvalues of a simple model of a scalar coupled to right-handed neutrinos (plotted in \cref{fig:eigenvalue_plot}) naturally generate a highly asymmetric potential for the scalar field as show in \cref{fig:imga,fig:imgb,fig:pica}. The contribution of this highly asymmetric potential grows with temperature and dominates the scalar field’s dynamics in the early universe.

Motivated by these general observations, we have studied a simple $1+1$ model of sterile neutrinos. We find that the altered potential of the scalar can both modify its cosmology, but also its present day dynamics. We find the dynamics of the scalar field are dominated by the relic potential in many regions of parameter space. Furthermore, based on the analysis of our simplified $1+1$ model, that viable parameter space requires scalar masses larger than $m_\phi> 10^{-9}~{\rm eV}$, relatively large couplings $g\gtrsim 10^{-5} $, and initial conditions where $\phi_0< \phi_c$. This region corresponds to the lower triangle in \cref{fig:parameter_space_left}.

These conclusions have immediate applications to searches for time-varying or distorted neutrino oscillations (DiNOs) \cite{Berlin:2016woy,Krnjaic:2017zlz,Brdar:2017kbt,Whisnant:2018oft,Huang:2018cwo,Dev:2020kgz,Losada:2021bxx,Huang:2022wmz,Dev:2022bae}. Furthermore, the temperature dependence of the relic potential, \cref{V_relic}, can lead to a temporarily modified equation of state for the scalar field. A large fraction of the available parameter space in \cref{fig:parameter_space_left} is shaded green, implying that the dark matter relic density would have a modified equation of state at some time between $z=1$ and $z\approx1100$. We consider this region to be disfavored, but not ruled out, in the absence of a detailed analysis. The modification of the equation state could be constrained using cosmological simulation tools such as CLASS \cite{lesgourgues:2011xxx}. 

One possibility we have not fully explored is the role of the high-temperature relic potential when $m_D \lesssim T$. We are most interested in the possibility of zero-crossings, and so $m_D$ dictates which approximation of $V_{\rm relic}$ in \cref{V_relic} is valid since the mass of the light neutrino approaches $m_L \sim m_D$ as the field approaches the zero-crossing point.
 The condition $m_D \gg T$, assumed in the main text, guarantees that the low-temperature limit of the relic potential is valid when computing the turning point closest to $\chi=0$.   As the temperature of the bath increases the coefficient of the relic potential changes from $\mu_\pm T^3$ to $\mu_\pm^2 T^2$ (see \cref{V_relic}). This change in behaviour can ``soften'' the potential barrier and allow relic neutrinos to undergo a transition from light to heavy states $\nu_L\rightarrow \nu_H$. Notice, in particular, that the energy density in the scalar field is proportional to $T^3$ and so at sufficiently high temperatures grows faster than $\mu_\pm T^2$. Whether or not this is cosmologically viable, and if any observational signatures exist, is an interesting question that could be studied further. 

Another interesting extension of this work would be to study the role of the relic potential for pseudo-Dirac neutrinos. These models have non-trivial cosmologies and can be probed down to very weak couplings \cite{Dev:2020kgz,Dev:2022bae}. The role of the relic potential would be qualitatively different in a pseudo-Dirac scenario since both eigenstates would be present in the early universe. The relic potential would become symmetric about $\chi=0$, and there could be an interesting interplay between the scalar's conservative dynamics in the potential, and its dissipative dynamics from $\nu_H\rightarrow \nu_L \phi$ decays. 

\pagebreak

In summary, we find that decoupled cosmic relics can substantially alter a scalar field’s dynamics whenever their mass depends\footnote{It also would be interesting to understand if density-dependent potentials could alter the cosmology of derivatively coupled scalars.} on the scalar's expectation value. This idea is well appreciated in the MaVNs literature, but seems less well studied in the context of ultralight dark mater (see \cite{Batell:2021ofv,Batell:2022qvr,Murgui:2023kig} however for analogs involving the thermal potential). Any future search for distorted neutrino oscillations should carefully consider the present-day dynamics of the scalar field. Our work also provides a well motivated model for temperature dependent and asymmetric potentials. The temperature dependent and highly asymmetric potentials sourced by the massive cosmic relic can lead to a modified equation of state for ULDM and spoil concordance between cosmological observations from different epochs.  

\section*{Acknowledgements}
We thank Mark Wise for collaboration during early stages of this work. We thank Dave McKeen, Linda Xu, Kim Berghaus, Matheus Hostert, Saurunas Verner, Yohei Ema, and especially Gordan Krnjaic Akshay Ghalsasi and Nashwan Sabti for helpful discussions. We are grateful to  Mark Wise, Pedro Machado, Clara Murgui, Akshay Ghalsasi, and Leonardo Badurina for feedback on early versions of this manuscript.

RP is supported by the Neutrino Theory Network under Award Number DEAC02-07CH11359. RP and ST are supported by the U.S. Department of Energy, Office of Science, Office of High Energy Physics under Award Number DE-SC0011632, and by the Walter Burke Institute for Theoretical Physics. 

\bibliographystyle{JHEP}
\bibliography{biblio}

\end{document}